\documentclass[aps, prb, twocolumn, superscriptaddress,showkeys
]{revtex4-2}
\usepackage{tikz}
\usepackage{tikz-feynman}
\usetikzlibrary{positioning}
\tikzfeynmanset{
	bare/.style={fermion},
	dressed/.style={fermion, very thick},
	selfenergy/.style={blob, minimum size=8pt}
}

\usepackage{graphicx}
\usepackage{dcolumn}
\usepackage{bm}
\usepackage{mathrsfs}
\usepackage{tikz}
\usepackage{hyperref}
\input epsf.tex
\usepackage{amssymb}
\usepackage{amsmath}
\usepackage{amsthm}
\usepackage{amsopn}
\usepackage{ulem}
\usepackage{cancel}
\usepackage{color}
\usepackage{float}
\usepackage{comment}
\usepackage{multirow}
\usepackage{physics}

\usepackage{bbold}

\hypersetup{
	colorlinks=true,       
	linkcolor=blue,          
	citecolor=blue,        
	filecolor=magenta,      
	urlcolor=blue           
}

\begin{document}
	
	\title{Effect of classical noises on the coherent population trapping based on the Green's function approach to the multiplicative stochastic processes}
	
	\author{M. Mirzaee}
	\affiliation{Laser and Plasma Research Institute, Shahid Beheshti University, Tehran 19839-69411, Iran}
	\author{B. Askari}
	\affiliation{Laser and Plasma Research Institute, Shahid Beheshti University, Tehran 19839-69411, Iran}
	\author{Ali Motazedifard}
	\email{alimotazedifard@ut.ac.ir}
	\affiliation{Department of Physics, University of Tehran, Tehran  14395‑547, Iran}
	\affiliation{Quantum Remote Sensing Lab, Quantum Metrology Group, Iranian Center for Quantum Technologies (ICQT), Tehran, Tehran 15998-14713, Iran}
	\affiliation{Institute of Quantum Science and Technology, University of Tehran, North-Kargar Street, Tehran 1439955961, Iran}
	\author{A. Dalafi}
	\email{corresponding author: a$_$dalafi@sbu.ac.ir}
	\affiliation{Laser and Plasma Research Institute, Shahid Beheshti University, Tehran 19839-69411, Iran}

	\begin{abstract}
		Inspired by the Green's function (GF) approach in quantum field theory (QFT) and many body physics, we have developed a mathematical formalism to investigate classical multiplicative stochastic processes. Based on this approach, the interacting GF of any dynamical system subjected to classical stochastic noises, which enter into the system equations of motion in a multiplicative way, can be obtained from the noninteracting (free of noise) GF through an infinite perturbative series which may converge to an exact closed form under special conditions. Using this formalism, we have studied the effects of classical noises of the driving laser on the coherent population trapping (CPT) which have a crucial role in the performance of CPT-based atomic clocks. If the stochastic variables are white (delta correlated) noises, the infinite series converges exactly to a closed form so that the exact time evolution of the system as well as its exact steady state is derived from it. Nevertheless, the presented formalism has the advantage to be extended to dynamical systems with multiplicative colored noises. We have shown that if the bandwidth of the colored noise is sufficiently larger than the system damping rate, the infinite series corresponding to the interacting GF can be approximated by the closed form. The presented formalism enables us to investigate all kinds of homogeneous and inhomogeneous broadening mechanisms on the CPT transmission resonance lineshape, including dephasing due to atomic collisions, power/Doppler broadenings, as well as the broadening mechanisms due to phase and amplitude fluctuations of driving laser and compared their destructive effect with each other. It should be emphasized that the presented formalism is applicable to any dynamical system with multiplicative stochastic noises and the CPT phenomenon is just a prototype for the application of the presented formalism in practice.
	\end{abstract}

	\maketitle

	\section{Introduction}\label{int}
	Coherent population trapping (CPT), which was first observed by Alzetta in 1976 \cite{Alzetta}, is one of the most important phenomena which has a crucial place in the field of atomic frequency standards and  precise timekeeping \cite{Vanier2005}. If the atoms are driven by two coherent optical fields so that a $\Lambda$-shape structure is formed in the atomic levels, the atoms lie in a so-called dark state through a quantum interference phenomenon called CPT under the two-photon resonance condition where the frequency difference of the two optical fields exactly matches the splitting between ground state hyperfine levels \cite{Zubairy Book}. In this way, the atomic medium becomes transparent in a very narrow frequency range when the frequency of either optical field is scanned around the two-photon resonance so that a CPT transmission resonance line appears. Since the formation of dark state is not possible unless the frequency difference of two laser fields matches the hyperfine ground state splitting with a good precision, it is not preferable to use two independent lasers due to their intrinsic phase/amplitude noises. For this reason, the two optical beams should be extracted from a single laser whose driving current is directly modulated using a tunable radio frequency (RF) oscillator. The function of a CPT-based atomic clock is based on locking the frequency of the RF oscillator to match the frequency difference between two hyperfine ground states of an alkali-metal atom like Rb or Cs \cite{Belcher2009}.
	
	The great advantage of a CPT-based atomic clock over other kinds, is the possibility of minimizing its physical size, power consumption, and cost with the maintenance of precision and stability to a good level. That is why almost all researches on chip-scale atomic clocks have been concentrated specially on the CPT phenomenon  \cite{Zhong Review, minicpt1, minicpt2,minicpt3}. Obviously, neither is it worth making use of expensive low noise lasers in such low cost miniaturized atomic clocks, nor does there exist any possibility of laser cooling and trapping of atoms. Therefore, the performance of such devices are mainly affected by the classical noises of their installed driving laser in addition to other broadening mechanisms like Doppler and collision broadenings which leads to the decoherence of dark state and deformation of the CPT transmission resonance line shape. Since the clock precision and stability are directly related to the width, height, and contrast \cite{Shah2007} of CPT transmission resonance line, the principal question that is raised in such cases is how much each broadening mechanism affects theses properties. 
	
	In order to answer this question, a rigorous mathematical formalism is needed to incorporate all the involved mechanisms into a single model. As is well-known, in the theory of open quantum systems \cite{Carmichael} the effect of spontaneous emission as the origin of natural broadening of the excited atomic levels as well as that of dephasing of hyperfine ground (metastable) sublevels due to atomics collisions are taken into account as two kinds of Lindblad terms in the master equation \cite{Carmichael} of a driven atomic ensemble interacting with the vacuum electromagnetic field of surrounding environment. Furthermore, the effect of Doppler broadening \cite{Loudon} can be included in the model in the framework of classical theory of random variables \cite{Papoulis} by considering the detuning of atomic transitions with the driving laser as a random variable which is a function of the random velocity of atom. In this way, the Doppler broadening contribution on the system steady state values can be calculated by taking an ensemble average over the density matrix elements obtained from the master equation using the corresponding probability distribution function \cite{Dop1 Vemuri,Dop2 Mompart,Dop3 Fan}.
	
	The phase and amplitude fluctuations of the driving laser are the origin of other kind of broadening mechanism which affects the CPT transmission resonance line effectively. Such fluctuations can be modeled as classical stochastic variables which enter in the master equation of the atom in a multiplicative way. The theory of multiplicative stochastic processes, introduced in the decade 1960 \cite{Redfield,Kubo 1962,Kubo 1963}, describes dynamical systems driven by some classical noises which appear in the system equation of motion as stochastic variable multiplied by the system variables. The mathematical foundation of the theory of multiplicative stochastic processes was extensively developed in 1972 by Fox \cite{Fox} who presented an exact approach to solve a stochastic Liouville equation with multiplicative white (delta correlated) Gaussian noises in the context of non-equilibrium statistical mechanics. Later, other mathematical methods based on path integral technique \cite{Wodkiewicz} and short-time iterative expansion approach to stochastic differential equation (SDE) \cite{Cook} were also proposed by other researchers. Using the theory of multiplicative stochastic processes, the effect of temporal fluctuations of the laser beam on several important quantum optical phenomena like resonance fluorescence, optical absorption, anti-bunching, and optical double resonance have been studied \cite{Agarwal1976,Agarwal1978,Agarwal1990,Lawande1987}. Furthermore, there are other theoretical approaches for modeling the driving pump fluctuations where either the stochastic field is considered as a quantum oscillator with a classical fluctuating frequency part \cite{Agarwal1978,Dalton1982} or it is considered as an infinite-temperature quantum reservoir which can be eliminated to obtain a master equation with an extra Lindblad super operator modeling the effect of stochastic field \cite{Swain1998,Kiely2021,Franco2019,Zeng2024,Danageozian}. Interestingly, the interface between theories of classical stochastic processes and quantum optics has been even more important in the modern quantum science because of the inevitable destructive effect of noises on the performance of emerging quantum technological devices \cite{Jiang2023}.
	
	In this paper, we have developed a mathematical formalism to investigate classical multiplicative stochastic processes based on the GF approach which has been previously used in QFT and quantum many body physics. In this approach, the GF of an interacting quantum system is obtained from the noninteracting one through an infinite perturbative series which may converge to an exact closed form under special conditions taking into account all the infinite orders of perturbation. Inspired from this idea, we have shown that such an approach can be generalized to the classical multiplicative stochastic processes. Using this formalism, we have studied the effects of classical noises of the driving laser on the CPT transmission resonance line shape which has a crucial role in the performance of a CPT-based atomic clock. It is shown that the system dynamics derived from the master equation takes the form of a system of homogeneous SDE in which the classical laser phase and amplitude noises enter in a multiplicative way. It should be emphasized that the presented formalism is applicable to any dynamical system (no matter being quantum or classical) with multiplicative stochastic noises and the CPT phenomenon is just a prototype for the application of the presented formalism in practice.

	\begin{figure*}[!t]
		\centering
		\includegraphics[trim=110mm 51mm 109mm 22mm, 
		clip,
		width=0.5\textwidth]{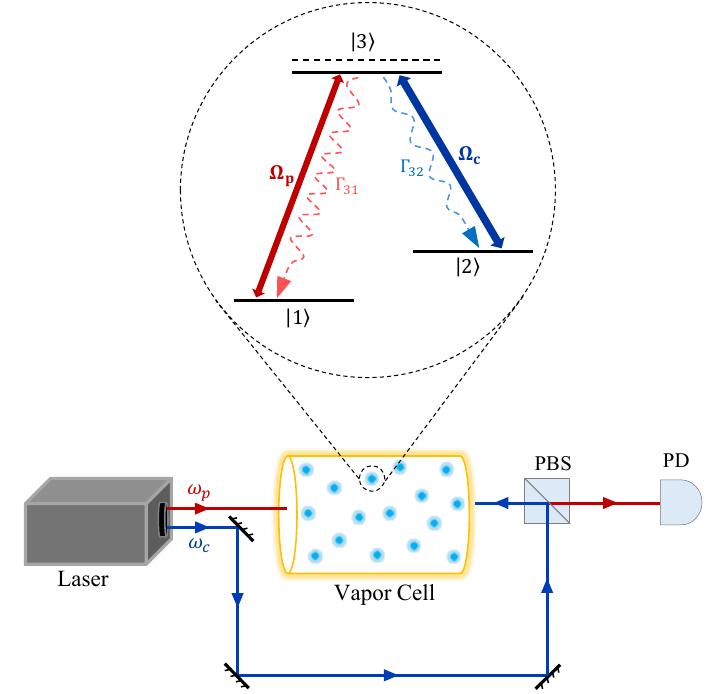}
		\caption{Schematic representation of the experimental setup for detection of the CPT transmission resonance line when a gas of alkali-metal atoms are subjected to the radiation of two beams extracted from the same laser source so that the atomic levels form a $\Lambda$-shape structure. The excited level $|3\rangle$ of each interacting atom has the spontaneous emission rates of $\Gamma_{31}$ and $\Gamma_{32}$ to the hyperfine ground state levels $|1\rangle$ and $|2\rangle$ while the $|1\rangle\to |2\rangle$ transition is dipole forbidden. The two coupling and probe beams with frequencies $\omega_c$ and $\omega_p$ are counterpropagating in order to neutralize the Doppler broadening as much as possible.}
		\label{figSchematic}
	\end{figure*}
	
	We show that a system of homogeneous multiplicative SDE can be solved using the perturbation theory so that it transforms to a system of linear inhomogeneous Langevin equations in every order of perturbation which can be solved based on the solutions to the previous order. In this way, the vector state of the system at any instant is obtained as a sum of different orders of perturbation. Then, by taking expectation values, the interacting (noisy) GF is derived as an infinite perturbative series whose each order is obtained in terms of noninteracting (noiseless) GF. If the stochastic variables are \textit{white} (delta correlated) noises, the infinite series converges \textit{exactly} to a \textit{closed form} so that the exact time evolution of the system as well as its exact steady state are derived from it. Nevertheless, the presented formalism in contrast to the other ones \cite{Fox,Wodkiewicz,Cook} which are applicable just to white (delta correlated) noises, has the advantage to be extended to dynamical systems with multiplicative \textit{colored noises}. We show that if the bandwidth of the colored noise is sufficiently larger than the system damping rate, the infinite series corresponding to the interacting GF can be \textit{approximated} by the closed form. In this way, the dynamics of the system and its steady state can be simply calculated again through the closed form of interacting GF with \textit{a good approximation}. Based on the presented formalism we have studied the effects of all kinds of homogeneous and inhomogeneous broadening mechanisms on the CPT transmission resonance lineshape, including dephasing due to atomic collisions, power and Doppler broadenings, as well as the broadening mechanism due to phase and amplitude fluctuations of driving laser and compared their destructive effect with each other.
	
	The paper has been structured as follows: In Sec.\ref{sysdis} the CPT phenomenon is introduced as a typical example whose dynamics can be expressed as a multiplicative SDE. In Sec.\ref{interactGformalism}, which forms the heart of paper, the mathematical formalism of GF approach to the multiplicative SDE is presented based on perturbation theory and the closed form of interacting GF is derived for both white and colored noises. Then in Sec.\ref{secDoppler}, the effect of Doppler broadening on the CPT is incorporated in the theoretical model in the framework of classical theory of random variables. The results corresponding to the effect of classical phase and amplitude noises of driving laser as well as that of other broadening mechanisms on the CPT transmission resonance line shape are presented in Sec.\ref{secResults}. Finally, the summary and conclusion is expressed in Sec.\ref{secSummary}. Furthermore, in the Appendix \ref{AppA}, we have studied Spectral analysis of stochastic electric fields having amplitude and phase fluctuations, and in the Appendix \ref{AppB} the detailed derivations related to the forth ordered GF have been presented.

	\section{system description and its Hamiltonian}\label{sysdis}
	The physical system that we are going to study is an ensemble of alkali-metal atoms exposed to the radiation of two beams extracted from the same laser source so that one of the beams, called the probe field with frequency $\omega_p$, is tuned to the transition $|1\rangle\to|3\rangle$ while the other one, called the coupling field with frequency $\omega_c$, is tuned to the transition $|2\rangle\to|3\rangle$. Under these conditions, the atomic levels form a $\Lambda$-shape structure as has been depicted in Fig.\ref{figSchematic}. The two mentioned beams can be obtained from the same laser source by modulating the laser phase and forming a frequency comb \cite{Belcher2009}. It is also assumed that the $|1\rangle\to |2\rangle$ transition is dipole forbidden and the hyperfine ground state $|2\rangle$ is a metastable level whose the only decoherence channel is due to atomic collisions with a dephasing rate of $\gamma_{12}$ while the excited level $|3\rangle$ has the spontaneous emission rates of $\Gamma_{31}$ an $\Gamma_{32}$ to the hyperfine ground state levels, $|1\rangle$ and $|2\rangle$, respectively. Since the two beams are assumed to be extracted from the same laser source, they have the same amplitude fluctuation $\delta\mathcal{E}(t)$, and the same fluctuating phase $\phi(t)$. In this way, the two field are modeled as the following classical stochastic variables	
	
	\begin{equation}\label{Epcdef}
		E_d(t) = (\mathcal{E}_{d} + \delta\mathcal{E}(t)) \, e^{-i(\omega_d t +\phi(t))},
	\end{equation}
	with $d=p$ or $c$ corresponding to probe or coupling, respectively. Here, $\mathcal{E}_{p(c)} $ is the constant amplitude of the probe (coupling) field, while $\delta\mathcal{E}(t)$ and $\phi(t)$ are real stochastic variables characterizing the field amplitude fluctuation and the stochastic phase, respectively, whose statistical properties have been explained in Appendix \ref{AppA}.
	
	The system Hamiltonian can be written as
	\begin{equation}\label{ExactH}
		\begin{aligned}
			\hat{H}(t) &=\,\hbar \omega_1 |1\rangle \langle 1|+\hbar \omega_2 |2\rangle \langle 2| + \hbar \omega_3 |3\rangle \langle 3| \vphantom{\int}\\
			&- \frac{\hbar}{2}\Big( (\Omega_{c}+ \delta\Omega(t)) e^{-i(\omega_c t+\phi(t))} |3\rangle \langle 2| \vphantom{\int}\\ 
			&+ (\Omega_{p} + \delta\Omega(t)) e^{-i(\omega_p t+\phi(t))} |3\rangle \langle 1|\Big) + \text{H.C.}
		\end{aligned}
	\end{equation}
	where $\omega_1, \omega_2$, and $\omega_3$ are the eigenfrequencies of the three atomic levels, and  $\Omega_{p} =\mathcal{E}_{p} d / \hbar$ and $\Omega_{c} =\mathcal{E}_{c} d / \hbar$ are the Rabi frequencies corresponding to the constant parts of the probe and coupling  field amplitudes with $d$ being the transition electronic dipole moment, while $\delta\Omega(t)$ is the Rabi frequency fluctuation originated from the fluctuating part of the field amplitude and is defined as $\delta\Omega (t) = \frac{d}{\hbar}\, \delta\mathcal{E}(t) $ with a correlation function of $\langle \delta\Omega(t) \, \delta\Omega(t') \rangle = C_{\Omega}(t - t')$, where the function $C_{\Omega}(t-t')$ is determined based on the amplitude correlation function as $C_{\Omega}(t - t') = \frac{d^2}{\hbar^2} C_{\mathcal{E}}(t - t')$. For more details see the Appendix.\ref{AppA}.
	
	Now, using the unitary transformation 
	\begin{equation}\label{UniTrans}
		\hat U(t) = e^{i(\omega_c - \omega_p) t|2\rangle \langle 2|}  \, e^{-i(\omega_p t + \phi(t))|3\rangle \langle 3|},
	\end{equation}	
	the Hamiltonian of Eq.\eqref{ExactH} is transformed as
	\begin{equation}\label{rotatedH}
		\begin{aligned}
			\hat{H}_{rf}(t) &= \hat U^{\dagger}(t)\hat H \hat U(t)-i U^{\dagger}(t)\dot{\hat U}(t) \\
			&=\hbar \delta |2\rangle \langle 2| + \hbar(\Delta_p - \dot{\phi}(t)) |3\rangle \langle 3| \\
			&\qquad - \frac{\hbar}{2} \Big( (\Omega_{c}+ \delta\Omega(t)) |3\rangle \langle 2| \\
			&\qquad \qquad \qquad + (\Omega_{p} + \delta\Omega(t)) |3\rangle \langle 1| \Big) + \text{H.C.},
		\end{aligned}
	\end{equation}
	where $\delta = \Delta_p - \Delta_c$ is the two photon detuning while $\Delta_p = \omega_{31} -\omega_p$ and $\Delta_c = \omega_{32} -\omega_c$ are the detunings of the probe and coupling frequencies from the corresponding atomic transition, respectively.
	The dynamics of laser-driven atomic system in the rotating frame is governed by the following master equation \cite{Carmichael}
	\begin{align}
		\frac{d \hat{\rho}(t)}{dt} =& \, \frac{1}{i\hbar} [\hat{H}_{rf}(t), \hat{\rho}(t)] + \mathcal{L}_{spon} (\hat{\rho}) + \mathcal{L}_{deph} (\hat{\rho})  \label{ExMast} \vphantom{\int}
	\end{align}
	where $\mathcal{L}_{spon} (\hat{\rho})$ characterizes the process of spontaneous emission form the excited level due to the interaction with the vacuum field and and $\mathcal{L}_{deph} (\hat{\rho})$ models the process of dephasing of the metastable level $|2\rangle$ due to elastic collisions in an atomic vapor, which are given by
	\begin{subequations}
		\begin{align}
			\mathcal{L}_{spon} (\hat{\rho}) = \, &\frac{\Gamma_{31}}{2} (2\hat{\sigma}_{13} \rho \hat{\sigma}_{31} - \hat{\sigma}_{33} \rho - \rho \hat{\sigma}_{33}) \notag \\
			+& \frac{\Gamma_{32}}{2} ( 2\hat{\sigma}_{23} \rho \hat{\sigma}_{32} - \hat{\sigma}_{33} \rho - \rho \hat{\sigma}_{33})  \label{LindSpon} \vphantom{\int},\\
			\mathcal{L}_{deph} (\hat{\rho}) = &\frac{\gamma_{21}}{2} ( 2\hat{\sigma}_{22} \rho \hat{\sigma}_{22} - \hat{\sigma}_{22} \rho - \rho \hat{\sigma}_{22}). \label{LindDeph}
		\end{align}
	\end{subequations}
	Here, $\Gamma_{31}$ and $\Gamma_{32}$ are the spontaneous emission rates from state $|3\rangle$ to states $|1\rangle$ and $|2\rangle$, respectively, and $\gamma_{21}$ is the dephasing rate of the metastable level $|2\rangle$. From the master equation of Eq.\eqref{ExMast}, one can obtain the equations of motion of density matrix elements for the atomic variables as
	\begin{subequations}
		\begin{gather}
			\dot{\rho}_{11}-\Gamma_{31} \, \rho_{33} + \frac{i}{2} \Omega_p (\rho_{13}-\rho_{31}) = \frac{i}{2} \delta\Omega(t) (\rho_{31}-\rho_{13})\label{rho11Ex}\vphantom{\int}\\
			\dot{\rho}_{21}+\frac{1}{2} (\gamma_{21} + 2i\delta) \rho_{21}-\frac{i}{2}\Omega_c\rho_{31}+\frac{i}{2}\Omega_p \, \rho_{23} \notag\\
			=\frac{i}{2}\delta\Omega(t)(\rho_{31}-\rho_{23})\label{rho21Ex}\vphantom{\int}\\
			\dot{\rho}_{22}-\Gamma_{32}\rho_{33}+\frac{i}{2}\Omega_c(\rho_{23}-\rho_{32})=\frac{i}{2}\delta\Omega(t)(\rho_{32}-\rho_{23})\label{rho22Ex}\vphantom{\int}\\
			\dot{\rho}_{31}+\frac{1}{2}(\gamma_{31}+2i\Delta_p)\rho_{31}-\frac{i}{2}\Omega_p(\rho_{11}-\rho_{33})-\frac{i}{2}\Omega_c\rho_{21}\notag\\ =\frac{i}{2}\delta\Omega(t)(\rho_{11}-\rho_{33}+\rho_{21})+i\dot{\phi}(t) \rho_{31}\label{rho31Ex}\vphantom{\int}\\
			\dot{\rho}_{32}+\frac{1}{2} (\gamma_{32} + 2i\Delta_c) \rho_{32} - \frac{i}{2} \Omega_c (\rho_{22}-\rho_{33}) - \frac{i}{2} \Omega_p \rho_{12} \notag\\
			=\frac{i}{2}\delta\Omega(t) (\rho_{12}+\rho_{22} -\rho_{33}) + i\dot{\phi}(t) \rho_{32} \label{rho32Ex}\vphantom{\int}\\
			\dot{\rho}_{33}+ \gamma \, \rho_{33} - \frac{i}{2} \Omega_p (\rho_{13}-\rho_{31}) - \frac{i}{2} \Omega_c (\rho_{23}-\rho_{32}) \notag\\
			=\frac{i}{2}\delta\Omega(t)(\rho_{13}-\rho_{31}+\rho_{23}-\rho_{32})\label{rho33Ex}		
		\end{gather}
	\end{subequations}
	where we have defined the total spontaneous emission rate out of the state $|3\rangle$ as $\gamma = \, \Gamma_{31}+\Gamma_{32}$. Furthermore, the coherence decay rate of the transition $|3\rangle\to|2\rangle$ is $\gamma_{32} = \, \Gamma_{31} + \Gamma_{32} +  \gamma_{21}$, while that of the transition $|3\rangle\to|1\rangle$ is $\gamma_{31}=\gamma$.

	Now, in order to convert the system of differential equations of the density matrix, i.e., Eqs.\eqref{rho11Ex}-\eqref{rho33Ex}, into the standard form of the classical Langevin equations, we introduce the following nine-component state vector
	\begin{equation}\label{R}
		\small{\boldsymbol{R}(t)} = [\, \rho_{11}(t),\, \rho_{12}(t),\, \rho_{13}(t),\, \ldots\,, \rho_{33}(t) \,]^{\text{T}},
	\end{equation}
	for description of system quantum state so that the set of Eqs.\eqref{rho11Ex}-\eqref{rho33Ex} can be written in the following form
	\begin{equation}\label{RDot}
		\bm{\dot{R}}(t) + \mathbf {M} \, \bm{R}(t) =  \bm{\Phi}(t) \bm{R}(t),
	\end{equation}
	in which the stochastic matrix $\bm{\Phi}$(t) has been defined as 
	\begin{equation}\label{PhiMatrixdef}
		\bm{\Phi}(t) =  \dot{\phi}(t) \mathbf{N} + \delta\Omega(t) \mathbf{L}.
	\end{equation}
	The Matrix $\mathbf {M}$ in Eq.\eqref{RDot} is the coefficient matrix which is given by
	\begin{widetext}
		\begin{equation}\label{M}
			\mathbf{M}=\frac{1}{2} \begin{pmatrix}
				0 & 0 & i\Omega_p & 0 & 0 & 0 & -i\Omega_p & 0 & -2\Gamma_{31} \\
				0 & \gamma_{21}-2i\delta & i\Omega_c & 0 & 0 & 0 & 0 & -i\Omega_p & 0 \\
				i\Omega_p & i\Omega_c & \gamma_{31}-2i\Delta_p & 0 & 0 & 0 & 0 & 0 & -i\Omega_p \\
				0 & 0 & 0 & \gamma_{21}+2i\delta & 0 & i\Omega_p & -i\Omega_c & 0 & 0 \\
				0 & 0 & 0 & 0 & 0 & i\Omega_c & 0 & -i\Omega_c & -2\Gamma_{32} \\
				0 & 0 & 0 & i\Omega_p & i\Omega_c & \gamma_{32}-2i\Delta_c & 0 & 0 & -i\Omega_c \\
				-i\Omega_p & 0 & 0 & -i\Omega_c & 0 & 0 & \gamma_{31}+2i\Delta_p & 0 & i\Omega_p \\
				0 & -i\Omega_p & 0 & 0 & -i\Omega_c & 0 & 0 & \gamma_{32}+2i\Delta_c & i\Omega_c \\
				0 & 0 & -i\Omega_p & 0 & 0 & -i\Omega_c & i\Omega_p & i\Omega_c & 2\gamma_{31}
			\end{pmatrix}
		\end{equation}	
	\end{widetext}
	Furthermore, the matrices $\mathbf{N}$ and $\mathbf{L}$ in the stochastic matrix of Eq.\eqref{PhiMatrixdef} are given by
	\begin{align}
		\mathbf{N}=i \,	\begin{pmatrix}
			0 & 0 & 0 & 0 & 0 & 0 & 0 & 0 & 0 \\
			0 & 0 & 0 & 0 & 0 & 0 & 0 & 0 & 0 \\
			0 & 0 &-1 & 0 & 0 & 0 & 0 & 0 & 0 \\
			0 & 0 & 0 & 0 & 0 & 0 & 0 & 0 & 0 \\
			0 & 0 & 0 & 0 & 0 & 0 & 0 & 0 & 0 \\
			0 & 0 & 0 & 0 & 0 &-1 & 0 & 0 & 0 \\
			0 & 0 & 0 & 0 & 0 & 0 & 1 & 0 & 0 \\
			0 & 0 & 0 & 0 & 0 & 0 & 0 & 1 & 0 \\
			0 & 0 & 0 & 0 & 0 & 0 & 0 & 0 & 0
		\end{pmatrix}, \notag \\[8pt]
		\mathbf{L}= \frac{i}{2} \begin{pmatrix}
			0  &  0 & -1 &  0 &  0 &  0 &  +1 &  0 & 0 \\
			0  &  0 & -1 &  0 &  0 &  0 &  0 &  +1 & 0 \\
			-1 & -1 &  0 &  0 &  0 &  0 &  0 &  0 & +1 \\
			0  &  0 &  0 &  0 &  0 & -1 &  +1 &  0 & 0 \\
			0  &  0 &  0 &  0 &  0 & -1 &  0 &  +1 & 0 \\
			0  &  0 &  0 & -1 & -1 &  0 &  0 &  0 & +1 \\
			+1  &  0 &  0 &  +1 &  0 &  0 &  0 &  0 & -1 \\
			0  &  +1 &  0 &  0 &  +1 &  0 &  0 &  0 & -1 \\
			0  &  0 &  +1 &  0 &  0 &  +1 & -1 & -1 & 0
		\end{pmatrix},	\label{Lmatrix}
	\end{align}
	It should be noted that Eq.\eqref{RDot} is a homogeneous differential equation in which the stochastic driving forces enter in a \textit{multiplicative} way. Such an equation is a multiplicative SDE \cite{Fox,Wodkiewicz}, which describes a \textit{nonequilibrium} dynamical system. The time evolution of the vector state $\bm R(t)$ is obtained by solving Eq.\eqref{RDot} with the initial condition of $\small{\boldsymbol{R}(t=0)} = \boldsymbol{R}_0$, where it can be considered as $ \boldsymbol{R}_0=[\, 1,\, 0,\, 0,\, \ldots\,, 0 \,]^{\text{T}}$, because before the beginning of interaction, the system has been in an equilibrium state where approximately most of the atomic population has been concentrated in the ground state. Nevertheless, in the long time compared with the relaxation time of the upper state (1/$\gamma$), the system reaches to a steady state that is independent of its initial state as will be shown in the next section.

	\section{mathematical formalism of GF approach to the multiplicative SDE}\label{interactGformalism}
	It should be noted that there are a plenty of physical phenomena which can be described by the multiplicative SDE of Eq.\eqref{RDot}, where the CPT phenomenon which was described in Sec.\ref{sysdis} is a prototype. In this section, we present a general mathematical formalism based on the GF approach to the multiplicative SDEs which can be applied to any stochastic dynamical equations with an equation of motion of Eq.\eqref{RDot}. Interestingly, the presented formalism is the classical stochastic version of the GF formalism in the literature of QFT and quantum many-body physics \cite{Flensberg}. Therefore, in the same manner that the GF of an interacting quantum field can be expanded perturbatively in terms of noninteracting GF and the self-energy, it will be shown that for a classical stochastic dynamical system such a formalism does function.
	
	\subsection{Solving multiplicative SDE based on perturbation theory}
	Here, we show how the multiplicative SDE of Eq.\eqref{RDot} can be solved using the \textit{perturbation} theory so that it transforms to a system of linear inhomogeneous Langevin equations in every order of perturbation which can be solved based on the solutions to the previous order. For this purpose we write Eq.\eqref{RDot} as
	\begin{equation}\label{RDotPer}
		\bm{\dot{R}}(t) + \mathbf {M} \, \bm{R}(t) = \, \varepsilon\bm\Phi'(t) \, \bm{R}(t),
	\end{equation}
	in which the stochastic matrix $\bm{\Phi}'(t)=\bm\Phi(t)/\varepsilon\bm$ has been redefined in terms of the rescaled stochastic variables $\dot{\phi}'(t) = \dot{\phi}(t)/\varepsilon$ and $\delta\Omega'(t) = \delta\Omega(t)/\varepsilon$. Here, $\varepsilon \ll 1$ has been included just for the purpose of a way to keep track of the different orders of perturbation in the calculations. 
	
	Now, if we expand the state vector $\mathbf{R}(t)$ in terms of different orders of perturbation as
	\begin{equation}\label{RPer}
		\mathbf{R}(t) = \sum_{n} \varepsilon^n \mathbf{R}^{(n)}(t),
	\end{equation}	
	and substitute it into Eq.\eqref{RDotPer}, the following linear inhomogeneous equations are derived in each order of perturbation with $n\ge 1$ by collecting terms of the same order of $\varepsilon$ 
	\begin{equation}\label{PerEqR}
		\bm{\dot{R}}^{(n)}(t) + \mathbf {M} \, \boldsymbol{R}^{(n)}(t) =  \bm{\Phi}'(t) \, \bm{R}^{(n-1)}(t),
	\end{equation}
	while a linear homogeneous equation is obtained for the zeroth order which is give by
	\begin{equation}\label{PerEqR-Zeroth}
		\bm{\dot{R}}^{(0)}(t) + \mathbf {M} \, \boldsymbol{R}^{(0)}(t) = 0.
	\end{equation}
	It should be emphasized that Eq.\eqref{PerEqR} is a system of linear inhomogeneous equation because $\boldsymbol{R}^{(n-1)}(t)$ on the right hand side, is a known state vector which is determined through the solution to the previous order of perturbation.
	In order to solve Eq.~\eqref{PerEqR}, we introduce the noninteracting retarded GF $\bm{G}(t, t')$, which satisfies the following equation
	\begin{equation}\label{eom_G0_t}
		\left( \frac{d}{dt} + \mathbf{M} \right) \bm{G}(t-t') = \delta(t - t') \mathbf{I},
	\end{equation}
	whose solution is given by
	\begin{equation}\label{G0_t}
		\bm{G}(t-t') = \theta(t - t') e^{-\bm{M}(t - t')},
	\end{equation}
	where $\theta(t - t')$ is the Heaviside step function. Using the noninteracting GF, the solution to Eq.~\eqref{PerEqR} can be expressed as 
	\begin{equation}\label{solordern}
		\bm{R}^{(n)}(t) = \int_0^t dt' \, \bm{G}(t-t') \bm{\Phi}'(t') \bm{R}^{(n-1)}(t')
	\end{equation}
	It is obvious that the solution to the zeroth order equation of Eq.(\ref{PerEqR-Zeroth}) is given by
	\begin{equation}\label{FinalR}
		\bm{R}^{(0)}(t) = \bm{G}(t) \bm{R}_0,
	\end{equation}
	where $ \bm{R}_0$ is a known initial state. Based on Eq.\eqref{solordern}, the solution to the first order perturbation is given by
	\begin{equation}\label{SolRn1}
		\bm{R}^{(1)}(t) =  \int_0^{t} dt' \, \bm{G}(t-t') \bm{\Phi}'(t')\bm{G}(t') \bm{R}_0,
	\end{equation}
	while the second order solution is calculated as
	\begin{align}
		\bm{R}^{(2)}(t) &=  \int_0^t dt' \, \bm{G}(t-t') \bm{\Phi}'(t') \bm{R}^{(1)}(t')\label{Rin-1}\nonumber\\
		&= \int_0^t dt' \int_0^{t'} dt'' \, \bm{G}(t- t') \bm{\Phi}'(t') \bm{G}(t'-t'') \nonumber\\[3.5pt]
		&\qquad \qquad \qquad \quad \quad \quad \quad
		\times \bm{\Phi}'(t'')	\bm{G}(t'') \bm{R}_0.
	\end{align}
	where in the second line, the solution to the first order has been substituted from Eq.\eqref{SolRn1}. Now, by taking the expectation value from both sides of Eq.(\ref{RPer}) we have
	\begin{align}\label{pertexpanRt}
		\langle\bm{R}(t)\rangle = \langle\bm{R}^{(0)}(t)\rangle + \varepsilon\, \langle\bm{R}^{(1)}(t)\rangle + \varepsilon^2\,\langle\bm{R}^{(2)}(t)\rangle + ...  
	\end{align}
	Using the fact that each perturbative term $\mathbf{R}^{(n)}(t)$ is generated by successive convolutions of the retarded GF with the noise correlations, we may factor out the initial vector $\mathbf{R}_0$ and rewrite the perturbative expansion of Eq.\eqref{pertexpanRt} in a compact form as 
	\begin{align}\label{Rt_intracting}
		\langle\bm{R}(t)\rangle = \bm{\mathcal{G}}(t) \bm{R}_0	
	\end{align}
	where 
	\begin{equation}\label{interactingGtexpan}
		\bm{\mathcal{G}}(t) = \bm{\mathcal{G}}^{(0)}(t) + \bm{\mathcal{G}}^{(1)}(t) + \bm{\mathcal{G}}^{(2)}(t) + \bm{\mathcal{G}}^{(3)}(t) + ... ,
	\end{equation}
	is the full GF of interacting system. In the following, we show that the interacting GF of Eq.\eqref{interactingGtexpan} has a closed form by considering all orders of perturbation up to infinity. In this way, the time evolution of the averaged state vector of the interacting (noisy) system can be calculated by Eq.\eqref{Rt_intracting} in terms of any arbitrary initial state $\bm R_0$.

	\subsection{derivation of the closed form of the interacting GF}
	The zeroth order term in the perturbative expansion of Eq.\eqref{interactingGtexpan} which is obtained as
	\begin{equation}
		\bm{\mathcal{G}}^{(0)}(t) = \theta(t) e^{-\bm{M}t}, \hspace{2cm}
	\end{equation}
	is nothing except for the noninteracting GF, while the first order term which is given by
	\begin{equation}
		\bm{\mathcal{G}}^{(1)}(t) = \int_0^{t} dt_1 \, \bm{G}(t-t_1) \langle\bm\Phi(t_1)\rangle\bm{G}(t_1) ,
	\end{equation}
	is zero  because $\langle\bm\Phi(t_1)\rangle = 0$ based on the definition of Eq.\eqref{PhiMatrixdef}. The second order term can be written as
	\begin{equation}\label{secondordGint}
		\bm{\mathcal{G}}^{(2)}(t) = \int_0^t dt_1 \int_0^{t_1} dt_2 \, 
		\bm{G}(t - t_1) \bm{\Sigma}(t_1-t_2)\bm{G}(t_2).
	\end{equation}
	in which the matrix of self-energy has been defined as
	\begin{align}\label{Secordselfenergy}
		\bm{\Sigma}(t_1-t_2) &= \Big\langle\bm\Phi(t_1)\bm{G}(t_1 - t_2)\bm\Phi(t_2)\Big\rangle \nonumber \\[5pt]
		&= \bm{\Sigma}_{\phi}(t_1-t_2) + \bm{\Sigma}_{\Omega}(t_1-t_2)
	\end{align}
	where the two matrices in the second line of Eq.\eqref{Secordselfenergy}, which are the contributions of the phase and amplitude noises to the total self-energy, are defined as
	\begin{subequations}
		\begin{align}
			\bm{\Sigma}_{\phi}(t_1-t_2)&\equiv C_{\phi}(t_1-t_2)
			\mathbf{N}\mathbf{G}(t_1-t_2)\mathbf{N},\label{SigmaPhit}\\[6pt]
			\bm{\Sigma}_{\Omega}(t_1-t_2)&\equiv C_{\Omega}(t_1-t_2)
			\mathbf{L}\mathbf{G}(t_1-t_2)\mathbf{L}\label{SigmaOmeghat}.
		\end{align}
	\end{subequations}
	Here, $C_{\phi}$ is always determined by Eq.\eqref{Cphifot} while $C_{\Omega}$ can be considered either as Eq.\eqref{CWwhit} if the amplitude fluctuation is a white noise, or as Eq.\eqref{CWcolor} if it is a colored noise.
	
	It should be noted that $\bm{\mathcal{G}}^{(3)}(t)$ and all other odd orders in the perturbative expansion of Eq.\eqref{interactingGtexpan} are zero because based on the Wick's theorem any odd moment is written as a sum of terms containing product of some second order moments and first order one which becomes zero if the stochastic noises have Gaussian statistics with zero average.
	
	In the same way, the forth order of the interacting GF is obtained as
	\begin{align}\label{G4tfirstform}
		\bm{\mathcal{G}}^{(4)}(t)
		&= \int_0^t dt_1\int_0^{t_1} dt_2\int_0^{t_2} dt_3\int_0^{t_3} dt_4\nonumber\\ 
		&\bm{G}(t-t_1)\Big\langle\bm{\Phi}(t_1)
		\bm{G}(t_1-t_2)\bm{\Phi}(t_2)
		\bm{G}(t_2-t_3)\nonumber\\ 
		&\quad \quad \quad \quad \quad \bm{\Phi}(t_3)\bm{G}(t_3-t_4)\bm{\Phi}(t_4)\Big\rangle \bm{G}(t_4).
	\end{align}
	In the Appendix \ref{AppB}, it has been shown that if the amplitude noise is delta correlated, then Eq.\eqref{G4tfirstform} is exactly reduced to the following equation
	\begin{align}\label{G4_t}
		\bm{\mathcal{G}}^{(4)}(t)
		&= \int_0^t dt_1\int_0^{t_1} dt_2\int_0^{t_2} dt_3\int_0^{t_3} dt_4 \nonumber\\[4pt]
		\times &\bm{G}(t-t_1)\bm{\Sigma}(t_1-t_2)
		\bm{G}(t_2-t_3)\bm{\Sigma}(t_3-t_4)
		\bm{G}(t_4).	
	\end{align}

	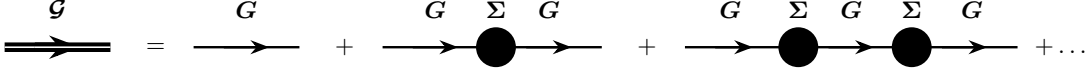
\begin{figure*}[!t]
		\centering
		\begin{tikzpicture}[>=Stealth, line width=0.9pt]
			
			\tikzset{
				G/.style={->},                     
				L/.style={-},                      
				Gdressed/.style={->,double,line width=1.5pt}, 
				Ldressed/.style={-,double,line width=1.5pt},  
				selfenergy/.style={circle, fill=black, minimum size=15pt, inner sep=0pt}
			}
			
			
			\draw[Gdressed] (0,0) -- (1,0);
			\draw[Ldressed] (0.4,0) -- (1.4,0);
			\node at (0.7,0.5) {$\bm{\mathcal{G}}$};
			
			\node at (2,0) {$=$};
			
			\draw[G] (2.5,0) -- (3.5,0);
			\draw[L] (2.9,0) -- (3.9,0);
			\node at (3.2,0.5) {$\bm{G}$};
			
			\node at (4.5,0) {$+$};
			
			\draw[G] (5,0) -- (6,0);
			\draw[L] (5.4,0) -- (6.4,0);
			\node at (5.7,0.5) {$\bm{G}$};
			
			\node[selfenergy] (sigma1) at (6.5,0) {};
			\node at (6.5,0.5) {$\bm{\Sigma}$};
			
			\draw[G] (6.5,0) -- (7.5,0);
			\draw[L] (6.9,0) -- (7.9,0);
			\node at (7.2,0.5) {$\bm{G}$};
			
			\node at (8.5,0) {$+$};
			
			\draw[G] (9,0) -- (9.9,0);
			\draw[L] (9.4,0) -- (10.4,0);
			\node at (9.6,0.5) {$\bm{G}$};
			
			\node[selfenergy] (sigma2) at (10.5,0) {};
			\node at (10.5,0.5) {$\bm{\Sigma}$};
			
			\draw[G] (10.5,0) -- (11.5,0);
			\draw[L] (10.9,0) -- (11.9,0);
			\node at (11.2,0.5) {$\bm{G}$};
			
			\node[selfenergy] (sigma3) at (12,0) {};
			\node at (12,0.5) {$\bm{\Sigma}$};
			
			\draw[G] (12,0) -- (13,0);
			\draw[L] (12.4,0) -- (13.4,0);
			\node at (12.8,0.5) {$\bm{G}$};
			
			\node at (14,0) {$+\dots$};
			
		\end{tikzpicture}
		\caption{Feynman diagram representation of the infinite series expansion of the interacting (noisy) GF. The interacting GF is demonstrated by the directional double line on the left hand side, while the noninteracting (noiseless) GF is represented by directional single lines and the self-energy contribution by solid black circles on the right hand side. The diagrams illustrate the repeated interaction processes appearing in the perturbative expansion of Eq.\eqref{Gintinfinitseries}. }
		\label{FeynmanD}
	\end{figure*}

	Nevertheless, it has been also explained in Appendix \ref{AppB} that if the amplitude fluctuation is a colored noise with an autocorrelation function of Eq.\eqref{CWcolor} whose frequency width ($\kappa$) is much larger than the damping rate of the upper atomic stater ($\gamma$), i.e., $\kappa\gg\gamma$, then Eq.\eqref{G4_t} is again valid with a good approximation.

	The fourth-order averaged contribution therefore takes the compact form of a two-fold convolution between retarded GFs and the corresponding second-order self-energies. It can be easily shown that all higher perturbation orders of GF (like the fourth order) in the time domain are in the form of a convolution integral. Therefore, they take the form of algebraic product in the frequency domain. For this reason from now on we follow our calculations in the frequency domain. By solving Eq.\eqref{eom_G0_t} in frequency space the noninteracting GF is obtained as 
	\begin{equation}\label{G0_omega}
		\widetilde{\bm{G}}(\omega) = \frac{1}{-i\omega \bm{I} + \bm{M}}.
	\end{equation}
	It should be noted that the second order GF of Eq.\eqref{secondordGint} can be written in the form of a convolution integral as 
	\begin{equation}
		\bm{\mathcal{G}}^{(2)}(t) = \int_{0}^{t}dt_{1} \bm{G}(t-t_{1})\bm{\mathcal{F}}(t_1),
	\end{equation} 
	in which $\bm{\mathcal{F}}(t_1)$ has been defined as
	\begin{align}\label{F_t}
		\bm{\mathcal{F}}(t_1) = \int_{0}^{t_{1}}dt_{2}\, \bm{\Sigma}(t_{1}-t_{2})\bm{G}(t_{2}).
	\end{align}

	Considering the Fourier transform as $\tilde{\bm{X}}(\omega)\equiv\int_{-\infty}^{\infty}dt e^{i\omega t}\bm{X}(t)$, the second order GF is transformed as
	\begin{align}
		\widetilde{\bm{\mathcal{G}}}^{(2)}(\omega)
		&=\int_{0}^{\infty}dt e^{i\omega t}\int_{0}^{t}dt_{1}
		\mathbf{G}(t-t_{1})\bm{\mathcal{F}}(t_1)\nonumber\\
		&=\int_{0}^{\infty}dt_{1}\int_{t_{1}}^{\infty}dt
		e^{i\omega t}\mathbf{G}(t-t_{1})\bm{\mathcal{F}}(t_1)
	\end{align}
	Here, the lower limits start from zero because there is a Heaviside theta function inside the GF $\bm{G}$.
	Now we consider the change of variables $t'=t-t_1$ and $t''= t_1$. Since the Jacobian of this change of variables equals to unity we have
	\begin{align}
		\tilde{\bm{\mathcal{G}}}^{(2)}(\omega)	&=\int_{0}^{\infty}dt''\int_{0}^{\infty} dt' \,
		e^{i\omega(t''+t')}\mathbf{G}(t')\bm{\mathcal{F}}(t'')\nonumber\\[5pt]
		&=\Big(\int_{0}^{\infty}dt' \, e^{i\omega t'}\mathbf{G}(t')\Big)
		\Big(\int_{0}^{\infty} dt'' \, e^{i\omega t''}\bm{\mathcal{F}}(t'')\Big)\nonumber\\[5pt]
		&=\ \widetilde{\mathbf{G}}(\omega) \widetilde{\bm{\mathcal{F}}}(\omega).
	\end{align}
	According to the convolution form of Eq.\eqref{F_t} the second order GF is read as 
	\begin{align}
		\widetilde{\bm{\mathcal{G}}}^{(2)}(\omega)	= \,\widetilde{\mathbf{G}}(\omega)\,\widetilde{\boldsymbol{\Sigma}}(\omega)\,\widetilde{\mathbf{G}}(\omega).
	\end{align} 
	Similarly, the Fourier transform of the fourth order retarded GF of Eq.\eqref{G4_t} is calculated as
	\begin{align}\label{G4omega}
		\widetilde{\bm{\mathcal{G}}}^{(4)}(\omega)	= \,\widetilde{\mathbf{G}}(\omega)\,\widetilde{\boldsymbol{\Sigma}}(\omega)\,\widetilde{\mathbf{G}}(\omega)\,
		\widetilde{\boldsymbol{\Sigma}}(\omega)\,\widetilde{\mathbf{G}}(\omega),
	\end{align}
	In the same manner, if the even orders with $n=6,8,...$ are calculated, the same pattern as that in the right hand side of Eq.\eqref{G4omega} is repeated. Therefor, the perturbative expansion of the interacting GF of Eq.\eqref{interactingGtexpan} is calculated as the following infinite series in Fourier space as 
	\begin{align}\label{Gintinfinitseries}
		\widetilde{\bm{\mathcal{G}}}(\omega) 
		=\widetilde{\bm{G}}(\omega)\bigg(1+ \,\widetilde{\bm{\Sigma}}(\omega)\widetilde{\bm{G}}(\omega) + \big( \widetilde{\bm{\Sigma}}(\omega)\widetilde{\bm{G}}(\omega)\big)^2 + ...\bigg).
	\end{align}

	The infinite series appeared inside the parentheses is a geometric series which has a closed form. So the interacting GF can be written as 
	\begin{align}\label{Gint_omega}
		\widetilde{\bm{\mathcal{G}}}(\omega) = \frac{1}{\widetilde{\bm{G}}^{-1}(\omega) - \widetilde{\bm{\Sigma}}(\omega)}
	\end{align}
	As is seen, the infinite perturbative series of the interacting GF converges to a closed form in the frequency space. Therefore, by taking the inverse Fourier transform of Eq.\eqref{Gint_omega} one can obtain the interacting GF in the time domain by which the state evolution of the interacting system is derived exactly from Eq.\eqref{Rt_intracting}.

	In order to show the repeated patterns appeared in the infinite series expansion of the interacting GF, we have represented in Fig.\ref{FeynmanD} the infinite series on the right hand side of Eq.\eqref{Gintinfinitseries} diagrammatically inspired from Feynman diagrams in the literature of QFT. Here, the interacting (noisy) GF has been shown by a directional double line on the left hand side of Fig.\ref{FeynmanD} while the nononteracting (noiseless) GF and the self-energy have been represented, respectively, by directional single lines and black solid circles on right hand side.

	It should be emphasized that although in the present work, Eq.\eqref{RDot} is the equation of motion corresponding to the density matrix time evolution of an atom exposed to a stochastic potential $V(t)$, but it can be also considered as a classical Langevin equation which describes a fully classical stochastic dynamical system whose classical state is described by the vector $\bm R(t)$.

	\subsection{White amplitude noise}\label{subwhitenoise}
	As a special case where  the amplitude fluctuation, just like the phase fluctuation, is a white noise with an autocorrelation function of Eq.\eqref{CWwhit} the Fourier transform of self-energy of Eq.\eqref{Secordselfenergy} is obtained as
	\begin{align}\label{self energy_omega}
		\widetilde{\bm{\Sigma}}(\omega) 
		= \int_{-\infty}^{\infty} d\tau \ e^{i\omega \tau}\ \Big(& 2\Gamma_{\phi} \delta(\tau) \bm{N} \bm{G}(\tau)\bm{N} \notag\\
		& \quad + 2\Gamma_{\Omega} \delta(\tau) \bm{L} \bm{G}(\tau)\bm{L} \Big).
	\end{align}
	On substitution of Eq.\eqref{G0_t} into Eq.\eqref{self energy_omega}, the self-energy takes the following simple form 
	\begin{align}\label{white self energy_omega}
		\widetilde{\bm{\Sigma}}(\omega) 
		= \Gamma_{\phi}\bm{N}^2 + \Gamma_{\Omega} \bm{L}^2,
	\end{align}
	which is a frequency independent matrix. In this way, the interacting GF of Eq.\eqref{Gint_omega} for the delta correlated noises can be obtained as 
	\begin{align}\label{Gint_omega_white}
		\widetilde{\bm{\mathcal{G}}}(\omega) = \frac{1}{-i\omega \bm{I} + \bm{M'}},
	\end{align}
	in which 
	\begin{equation}\label{ModifiedM}
		\bm{M'} = \bm{M} - \Gamma_{\phi}\bm{N}^2 - \Gamma_{\Omega} \bm{L}^2.
	\end{equation}
	It shows that when the system is exposed to white noises, it behaves effectively as a noninteracting system with a modified coefficient matrix $\bm{M'}$. This result is in complete coincidence with the results of previous papers which had been obtained by different methods \cite{Fox,Cook}. Nevertheless, our formalism has the great \textit{advantage} that can be generalized to dynamical system exposed to colored noises as will be shown in the next subsection.

	The important point that should be noted is that based on Eq.\eqref{ModifiedM} and the form of matrix $\bm N$, the \textit{phase noise} manifests itself as the \textit{modification of the decoherence decay rates} of transitions $|3\rangle\to|1\rangle$ and $|3\rangle\to|2\rangle$, respectively, as $\gamma_{31}\to\gamma_{31}+2\Gamma_{\phi}$ and $\gamma_{32}\to\gamma_{32}+2\Gamma_{\phi}$ without changing the dephasing rate $\gamma_{12}$. However, the amplitude noise has much more complicated effects on the system decoherences because of the complicated form of Matrix $\bm L$.

	In order to study the system dynamics we need to have the GF in the time domain. For this purpose, we can take the inverse Fourier transform of Eq.\eqref{Gint_omega_white} to obtain the interacting GF of the system as
	\begin{equation}\label{G}
		\bm{\mathcal{G}}(t) = \theta(t)\ e^{-\bm{M}' t},
	\end{equation}
	which is just like the noninteracting GF of Eq.\eqref{G0_t} but with a modified coefficient matrix. Now the time evolution of the system state can be calculated using Eq.\eqref{Rt_intracting}. It should be noted that Eq.\eqref{G} is the GF of the following equation of motion 
	\begin{equation}
		\langle\bm{\dot{R}}(t)\rangle + \mathbf {M'} \, \langle\bm{R}(t)\rangle = 0.
	\end{equation}
	Such a dynamical system has a steady state solution which is obtained by the condition of $\langle\bm{\dot{R}}(t)\rangle=0$ which leads to the following equation
	\begin{equation}\label{StStWhiteNoise}
		\mathbf {M'} \, \langle\bm{R}_{ss}\rangle = 0.
	\end{equation}
	As is seen, the steady state of the system, which is derived from Eq.\eqref{StStWhiteNoise}, is independent of the initial state.

	\subsection{Colored amplitude noise}
	Here, we show that our formalism is not restricted just to the white amplitude noise. In general, for an amplitude noise with an autocorrelation function of $C_\Omega(\tau)$, the self-energy in the frequency space is calculated as
	\begin{align}\label{colored self energy_omega1}
		\widetilde{\bm{\Sigma}}(\omega) 
		= \int_{-\infty}^{\infty} d\tau \ e^{i\omega \tau}\ \Big(& 2\Gamma_{\phi} \delta(\tau) \bm{N} \bm{G}(\tau)\bm{N} \notag\\
		&\quad \quad + C_{\Omega}(\tau) \bm{L} \bm{G}(\tau)\bm{L} \Big),
	\end{align}
	which for an autocorrelation function in the form of Eq.\eqref{CWcolor}, it is simplified as
	\begin{align}\label{colored self energy_omega2}
		\widetilde{\bm{\Sigma}}(\omega) 
		= \Gamma_{\phi}\bm{N}^2 + \kappa \Gamma_{\Omega}\bm{L} \widetilde{\bm{G}}(\omega+i\kappa)\bm{L}.
	\end{align}
	As is seen, in spite of Eq.\eqref{white self energy_omega} which was frequency independent, Eq.\eqref{colored self energy_omega2} is a function of frequency $\omega$. Now, on substitution of Eq.\eqref{colored self energy_omega2} into Eq.\eqref{Gint_omega}, the interacting GF for colored amplitude noise in the frequency space is obtained.
	
	As is evident, the inverse Fourier transform of Eq.\eqref{Gint_omega} can be calculated numerically in the cases where the analytical solution is not possible. In this way, the time evolution of the system can be derived at all of the times. However, since we are interested in steady state solution, it is not necessary to study the transitive behavior of the system before reaching to the steady state. We will show that in the asymptotic limit of $t\to\infty$ it suffices to study the inverse Fourier transform of Eq.\eqref{Gint_omega} just at its zero pole. It should be noted that the poles of GF are complex variables whose real and imaginary parts indicate the frequencies and damping rates of the system normal modes. The zero pole is the only normal mode of the system which remains in the limit of $t\to\infty$ because all other modes which have nonzero damping rates are dissipated in the long time.
	
	In order to illustrate this matter more rigorously, Let us start with the inverse Fourier transform of Eq.\eqref{Gint_omega} which gives us the time domain GF as
	\begin{align}\label{Gomega_inv}
		\bm{\mathcal{G}}(t) = \int_{-\infty}^{\infty} \frac{d\omega}{2\pi} \frac{e^{-i\omega t}}{-i\omega\bm{I} + \bm{M} - \widetilde{\bm{\Sigma}}(\omega)},
	\end{align}
	in which Eq.\eqref{G0_omega} has been substituted for the noninteracting GF. The important point in calculation of an inverse Fourier integral like Eq.\eqref{Gomega_inv} in the asymptotic limit of $t\to\infty$ is that the function $\widetilde{\bm{\mathcal{G}}}(\omega)$ is a slow varying function in comparison with $e^{-i\omega t}$, because it oscillates very fast in the frequency space for large values of $t$  (the larger $t$ the faster oscillation of $e^{-i\omega t}$ in terms of $\omega$). Therefore the integral is zero at all frequencies except for those in which the denominator is zero, i.e., at the poles of GF which are introduced by the complex valued numbers $\omega_k$'s. In this way, $\bm{\mathcal{G}}(t)$ can be broken into a sum of several integrals as
	\begin{align}\label{Gomega_inv_expad}
		\bm{\mathcal{G}}(t) 
		&= \sum_k\int_{-\infty}^{\infty} \frac{d\omega}{2\pi}  \frac{e^{-i\omega t}}{[-i\omega\bm{I} + \bm{M} - \widetilde{\bm{\Sigma}}(\omega)]_{\omega \approx \omega_k}}.
	\end{align}
	It should be noted that each of the integrals inside the summation have nonzero values just around its corresponding pole of $\omega_k$. 
	The denominator of the integrand of Eq.\eqref{Gomega_inv_expad} can be expanded up to the first order of Taylor expansion as
	\begin{align}
		\bm{\mathcal{G}}(t)=\sum_k &\int_{-\infty}^{\infty} \frac{d\omega}{2\pi} \notag\\
		\times &\frac{e^{-i\omega t}}{-i\omega\bm{I} + \bm{M} - \widetilde{\bm{\Sigma}}(\omega_k) - (\omega - \omega_k)\partial_\omega \widetilde{\bm{\Sigma}}(\omega_k)}
	\end{align}
	where $\partial_\omega\widetilde{\bm{\Sigma}}(\omega_k)=\partial_\omega\widetilde{\bm{\Sigma}}(\omega)|_{\omega = \omega_k}$. 
	
	A system whose dynamics is governed by Eq.~\eqref{RDot} with a constant coefficient matrix $\bm M$ generally exhibits a transient time-dependent behavior before relaxing toward a steady state in the limit $t\to\infty$. In the presence of colored noise, the intermediate dynamics can be obtained from Eq.~\eqref{Gomega_inv_expad}. However, if one is interested only in the asymptotic regime, it is sufficient to retain the contribution associated with the zero-frequency pole $\omega_k=0$, since all nonzero poles are exponentially damped at large times. In this way, the interacting GF in the long-time limit can be written as
	\begin{align}\label{Gt_Z}
		\bm{\mathcal{G}}(t)
		=\bm{Z}\int_{-\infty}^{\infty} \frac{d\omega}{2\pi}
		\frac{e^{-i\omega t}}
		{-i\omega \bm{I} + \bm{Z}(\bm{M} - \widetilde{\bm{\Sigma}}(0))},
	\end{align}
	where the matrix $\bm Z$ has been defined as
	\begin{equation}
		\bm{Z}=
		\left(\bm{I}-i\partial_\omega\widetilde{\bm{\Sigma}}(0)\right)^{-1}.
	\end{equation}
	To evaluate Eq.~\eqref{Gt_Z}, we consider the matrix
	\begin{equation}
		\bm{M}_{\rm eff}
		=
		\bm{Z}\left(\bm M-\widetilde{\bm\Sigma}(0)\right),
	\end{equation}
	as the modified version of coefficient matrix $\bm M$ in the presence of colored noise and diagonalize it. For this purpose, let $\bm r_n$ and $\bm \ell_n$ denote the corresponding right and left eigenvectors of $\bm{M}_{\rm eff}$ with eigenvalues $\lambda_n$ satisfying
	\begin{subequations}
		\begin{align}
			\bm{M}_{\rm eff}\bm r_n
			&=
			\lambda_n \bm r_n,\\[5pt]
			\bm \ell_n^{T}\bm{M}_{\rm eff}
			&=
			\lambda_n \bm \ell_n^{T},
		\end{align}
	\end{subequations}
	while they also satisfy the following orthonormality and completeness relations
	\begin{subequations}
		\begin{gather}
			\bm \ell_m^{T}\bm r_n =\delta_{mn},\\[5pt]
			\sum_n \bm r_n \bm \ell_n^{T} =\bm I,\label{completeness}
		\end{gather}
	\end{subequations}
	The mentioned eigenvectors form a kind of basis where the effective coefficient matrix can be written in terms of them as the following form
	\begin{equation}
		\bm{M}_{\rm eff}
		=
		\sum_n \lambda_n \bm r_n \bm \ell_n^{T}.
	\end{equation}
	Using the completeness relation of Eq.\eqref{completeness}, the resolvent inside Eq.~\eqref{Gt_Z} can be decomposed as
	\begin{equation}
		\left[-i\omega \bm I+\bm{M}_{\rm eff}\right]^{-1}
		=
		\sum_n
		\frac{\bm r_n \bm \ell_n^{T}}
		{-i\omega+\lambda_n}.
	\end{equation}
	Substituting this expression into Eq.~\eqref{Gt_Z} yields
	\begin{align}
		\bm{\mathcal G}(t)
		&=
		\bm Z
		\sum_n
		\bm r_n \bm \ell_n^{T}
		\int_{-\infty}^{\infty}
		\frac{d\omega}{2\pi}
		\frac{e^{-i\omega t}}
		{-i\omega+\lambda_n}.
	\end{align}
	The remaining frequency integral can be evaluated using the Cauchy integral formula, giving
	\begin{align}
		\bm{\mathcal G}(t)
		&=
		\bm Z
		\sum_n
		e^{-\lambda_n t}
		\bm r_n \bm \ell_n^{T}.
	\end{align}
	Since $\mathrm{Re}[\lambda_n]>0$ for all nonstationary modes, every contribution except the zero mode decays exponentially with time. Consequently, in the asymptotic limit $t\to\infty$, only the eigenmode corresponding to $\lambda_0=0$ survives, yielding
	\begin{align}\label{Gt_inf}
		\bm{\mathcal G}(\infty)
		=
		\bm Z \bm r_0 \bm \ell_0^{T}.
	\end{align}
	If the zero mode is unique, then the steady state is unique as well, and the asymptotic dynamics becomes
	\begin{align}\label{StStColoredN}
		\langle \bm R_{\rm ss}\rangle
		=
		\bm{\mathcal G}(\infty)\bm{R_0}
	\end{align}
	Based on Eqs.\eqref{Gt_inf} and \eqref{StStColoredN}, the only component of the initial state vector $\bm R_0$ which is in the direction of $\bm r_0$ is remained 
	so that the renormalized right zero mode $\bm Z\bm r_0$ acts as the global attractor (steady state) of the dynamics which is independent of the initial state. Hence, the system loses memory of all transient components of the initial condition in the long-time limit~\cite{breuer2002theory}.

	\section{Effect of Doppler broadening on the CPT}\label{secDoppler}
	
	As already stated, the formalism outlined in the previous section, is valid only for static atoms. Now, we want to take into account the random atomic motions due to the thermal agitation in our formalism which leads to the Doppler broadening. In atomic vapors at thermal equilibrium at a finite temperature, atoms are not stationary but move with random velocities. At temperature $T$, the probability density function corresponding to the random variable $\bar v$ (the random atomic velocity along the laser beam direction, say $z$) follows the Maxwell-Boltzmann distribution \cite{Loudon}
	\begin{equation}\label{Maxwell–BoltzmannDist}
		f_{\bar v}(\bar v) =\sqrt{\frac{m}{2\pi k_B T}} \,\exp(-\frac{m_A \bar v^2}{2k_{B} T}),
	\end{equation}
	where $m_A$ is the atomic mass and $k_B$ is the Boltzmann constant. Note that in this section we denote the random variables with over bar. Because of the random motions, each atom experiences a random Doppler shift in the received frequency of the light beam it interacts with. For a single atom, moving with a velocity $\bar v$ along the $z$ axis, the probe frequency $\bar\omega_p$, which is seen by the atom, is also a random variable as a function of the random variable $\bar v$, which is approximated as
	\begin{equation}\label{omegap-dop}
		\bar\omega_p\approx\omega_p(1 \pm \frac{\bar v}{c}),
	\end{equation}
	in the non-relativistic limit of $v\ll c$, where $c$ is the speed of light. The upper sign corresponds to the situation where the atom moves toward the probe beam, while the lower sign corresponds to the situation where the atom moves away from it. This process leads to the Doppler broadening, which is an inhomogeneous broadening mechanism of the spectral line \cite{Loudon}.
	
	Similarly, since the coupling beam propagates in the same direction as that of the probe beam (although their orientations could be opposite), the coupling frequency $\bar\omega_c$, which is seen by the atom, is a random variable as a function of the random variable $\bar v$, which is approximated as
	\begin{equation}\label{omegac-dop}
		\bar\omega_c\approx\omega_c(1 \pm \frac{\bar v}{c}).
	\end{equation}
	Again, the upper sign corresponds to the situation where the atom moves toward the coupling beam, while the lower sign corresponds to the situation where the atom moves away from it. Therefore, the detunings of probe and coupling beams from their respective transition frequencies, as seen by the moving atom, are  denoted by
	\begin{subequations}
		\begin{eqnarray}
			\bar\Delta_p = \omega_{31}-\bar\omega_p=\Delta_p\mp \frac{\omega_p}{c} \bar v,\label{deltap-dop}\vphantom{\int}\\
			\bar\Delta_c = \omega_{32}-\bar\omega_c=\Delta_c\mp \frac{\omega_c}{c} \bar v.\label{deltac-dop}
		\end{eqnarray}
	\end{subequations}
	It should be emphasized that, in spite of $\Delta_c$ and $\Delta_p$ which are fixed parameters, $\bar\Delta_c$ and $\bar\Delta_p$ are random variables because they are functions of the random variable $\bar v$. Based on Eqs.\eqref{deltap-dop} and \eqref{deltac-dop}, we can rewrite $\bar\Delta_c$ in terms of $\bar\Delta_p$ as
	\begin{equation}\label{deltac-dop2}
		\bar\Delta_c = \Delta_c \pm \frac{\omega_c}{\omega_p}(\bar\Delta_p-\Delta_p),
	\end{equation}
	where, the plus sign corresponds to the situation where the probe and coupling beams are counterpropagating, while the minus sign corresponds to the situation where they are copropagating.
	
	Based on the classical theory of random variables \cite{Papoulis} if a random variable $\bar y$ is a linear function of another random variable $\bar x$ as $\bar y = a\bar x + b$, then the probability density function $f_{\bar y}(\bar y)$ is obtained in terms of the probability density function $f_{\bar x}(\bar x)$ as \cite{Papoulis}
	\begin{equation}\label{funofrandv}
		f_{\bar y}(\bar y)=\frac{1}{|a|} f_{\bar x}\Big(\frac{\bar y-b}{a}\Big).
	\end{equation}
	As is evident from Eq.\eqref{deltap-dop}, $\bar\Delta_p$ is a linear function of $\bar v$ with $a=\mp\frac{\omega_p}{c}$ and $b=\Delta_p$. Therefore, the probability density function of random variable $\bar\Delta_p$ can be obtained from Eq.\eqref{Maxwell–BoltzmannDist} using  Eq.\eqref{funofrandv} as
	\begin{equation}\label{dopDist}
		f_{\bar\Delta_p}(\bar\Delta_p) = \frac{1}{\sqrt{2\pi D^2}} \, e^{-\frac{(\bar\Delta_p -\Delta_p)^2}{2 D^2}}
	\end{equation}
	where $D$, i.e., the standard deviation of $\bar\Delta_p$, which is called the Doppler width, is given by
	\begin{equation}\label{dopWidth}
		D = \frac{\omega_p}{c} \sqrt{\frac{k_B T}{m}}.
	\end{equation}
	It means that $\bar\Delta_p \sim N(\Delta_p,D^2)$ is a normal random variable with a Gaussian density function of Eq.\eqref{dopDist} whose average and variance are, respectively, $\Delta_{p}$ and $D^2$.
	
	Now, in order to incorporate the effect of Doppler broadening into our formalism, it suffices to substitute the random variable $\bar\Delta_{p}$ for the fixed parameter $\Delta_{p}$ in the Matrix of Eq.\eqref{M}. Furthermore, since $\bar\Delta_c$ can be expressed in terms of $\bar\Delta_{p}$ based on Eq.\eqref{deltac-dop2}, we can consider $\bar\Delta_{p}$ as the only independent random variable with the probability density function of Eq.\eqref{dopDist}. Therefore, the coefficient Matrix of Eq.\eqref{M} will be a function of the random variable $\bar\Delta_{p}$ as $\bar{\bm M} (\bar\Delta_{p})$. In this way, we can construct an ensemble of steady states $\langle\bar{\bm R}_{ss}(\bar\Delta_p)\rangle$ in terms of different values of $\bar\Delta_p$ either based on Eq.\eqref{StStWhiteNoise} in the case of white amplitude noise or based on Eq.\eqref{StStColoredN} in the case of colored amplitude noise. Then, using the probability distribution function of Eq.\eqref{dopDist}, we can obtain the ensemble average of the random vector $\langle\bar{\bm R}_{ss}(\bar\Delta_p)\rangle$ in the steady state as follows
	\begin{equation}\label{rhoDop}
		\langle\langle\bar{\bm R}_{ss}\rangle\rangle = \int_{-\infty}^{+\infty}\bm \langle\bar{\bm R}_{ss}(\bar\Delta_p)\rangle \, f_{\bar\Delta_p}(\bar\Delta_p) \, d\bar\Delta_p.
	\end{equation}
	The double average sign on the left hand side of Eq.\eqref{rhoDop} means that we have taken two kinds of expectation values where the former corresponds to stochastic noises and the latter to the random variable $\bar\Delta_{p}$. Finally, it should be reminded that for the counterpropagating beams the plus sign is used in Eq.\eqref{deltac-dop2} while for copropagating the minus sign is used.

	\section{Numerical results and experimental discussion}\label{secResults}
	
	\begin{figure*}[!t]
		\centering
		\includegraphics[width=11cm]{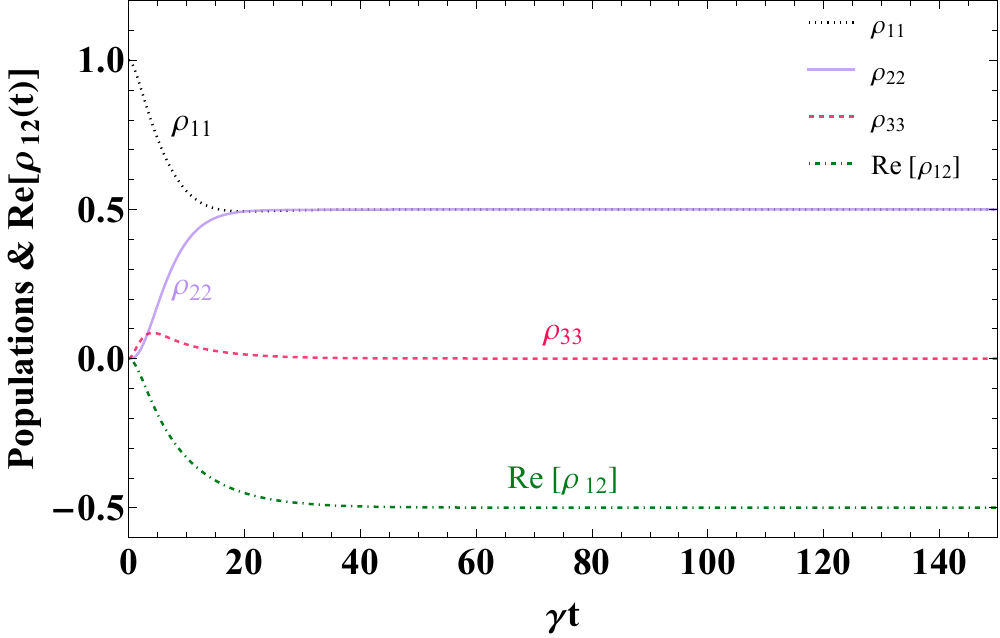}
		\caption{Time evolution of the populations of energy levels as well as the coherence between ground state hyperfine levels in the absence of noises and broadening mechanisms with the initial condition where the atom is its ground state assuming $\gamma_{12}=0$. The lasers parameters have been considered as $\Omega_{c}=\Omega_{p}=0.4\gamma$, $\Delta_c=0$, and $\delta=0$.}
		\label{figDynamicsDarkState}
	\end{figure*}
	
	In the first part of this section, we study the effects of each broadening mechanism separately on the CPT phenomenon while the other ones are absent. Then, in the next part we will show their collective effects in a real experimental situation which is applicable for CPT-based atomic clocks.

	\subsection{numerical results}
	In this subsection we study the effects of all broadening mechanisms as well as classical noises of the driving laser on the CPT phenomenon using the theoretical model presented in the previous sections. First of all, we study the dynamics of the system in the absence of any kind of noises or broadening mechanisms to show how the system evolves in time until it reaches to a steady state known as the dark state. For this purpose, it suffices to use the zero order solution of Eq.\eqref{FinalR} using the noninteracting GF of Eq.\eqref{G0_t} with the initial condition of $ \boldsymbol{R}_0=[\, 1,\, 0,\, 0,\, \ldots\,, 0 \,]^{\text{T}}$. 
	
	In order to determine the coefficient matrix of Eq.\eqref{M} numerically, we normalize all the frequencies in terms of the natural linewidth of the upper state $|3\rangle$ defined as $\gamma=\Gamma_{31}+\Gamma_{32}$. Furthermore, in our calculations we fix $\Gamma_{31}=\Gamma_{32}=\gamma/2$ for simplicity. It should be also reminded that although the two transitions $|1\rangle\to|3\rangle$ and $|2\rangle\to|3\rangle$ can have different dipole moments, in Sec. \ref{sysdis} we have considered the same dipole moment $d$ for both of them in the definition of Rabi frequencies $\Omega_{c(p)}$ and the fluctuation $\delta\Omega(t)$. Both of these assumptions are exact especially for the $D_1$ line of $^{87}\text{Rb}$ ($	5 ^{2}S_{1/2} \rightarrow 5 ^{2}P_{1/2}$) \cite{Rb87Steck}. Nevertheless, in the general cases where the two hyperfine transitions may have different dipole moments, it suffices to substitute the ratio of two dipole moments for those elements with unit absolute value in the matrix $\bm L$ of Eq.\eqref{Lmatrix} while the whole body of formalism remains unchanged.
	
	As is well known in the context of dynamical systems, the most general solution to the equations of motion of a linear dynamical system is a linear combination of its normal modes. If one of the normal modes has zero frequency and zero damping rate, the system goes to a steady state in the long time which is nothing except for that zero normal mode. It is because of the fact that all other normal modes disappear at times much longer than the inverse of damping rate. The phenomenon of CPT in the absence of stochastic noises is one of the most interesting examples of such linear dynamical systems, because Eq.\eqref{PerEqR-Zeroth}, which is the system equation of motion in the absence of stochastic noises, is linear. In order to see the phenomenon of CPT, it suffices to let the coupling and probe lasers have equal intensities while adjust the coupling laser detuning to zero $(\Delta_c=0)$. Under these conditions, among the three system normal modes one of them will have zero eigenvalue if the metastable transition $|1\rangle\to|2\rangle$ has zero dephasing rate ($\gamma_{21}=0$) while the two others have nonzero frequencies and damping rate. In this case, the zero normal mode, called the dark state $|D\rangle$, is generated at $\delta=0$ in the steady state which is a coherent combination as
	\begin{equation}\label{DarkS}
		|D\rangle=\frac{1}{\sqrt{2}}(|1\rangle - |2\rangle).
	\end{equation}
	
	\begin{figure*}[!t]
		\centering
		\includegraphics[width=8cm]{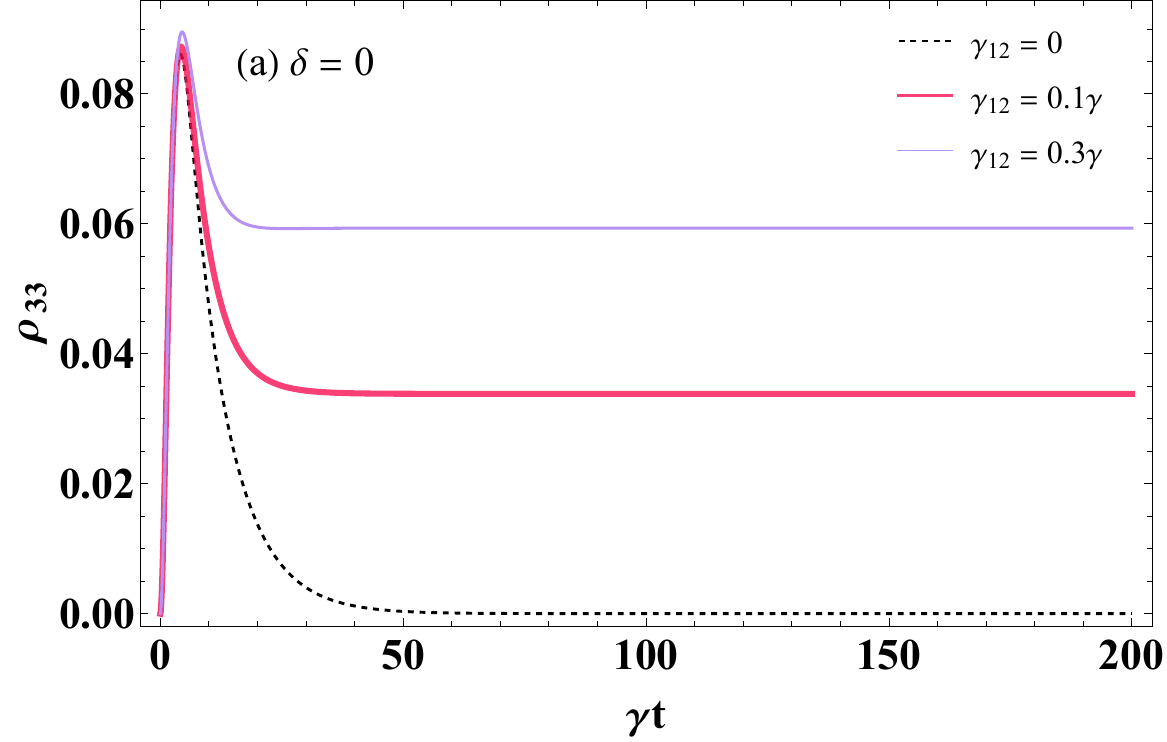}
		\includegraphics[width=8cm]{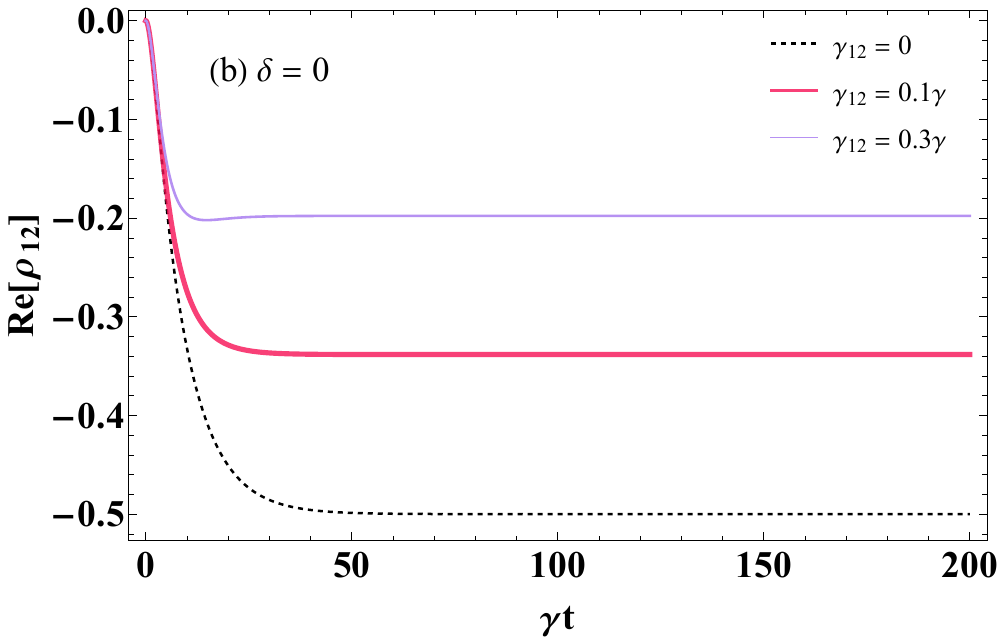}
		\includegraphics[width=8cm]{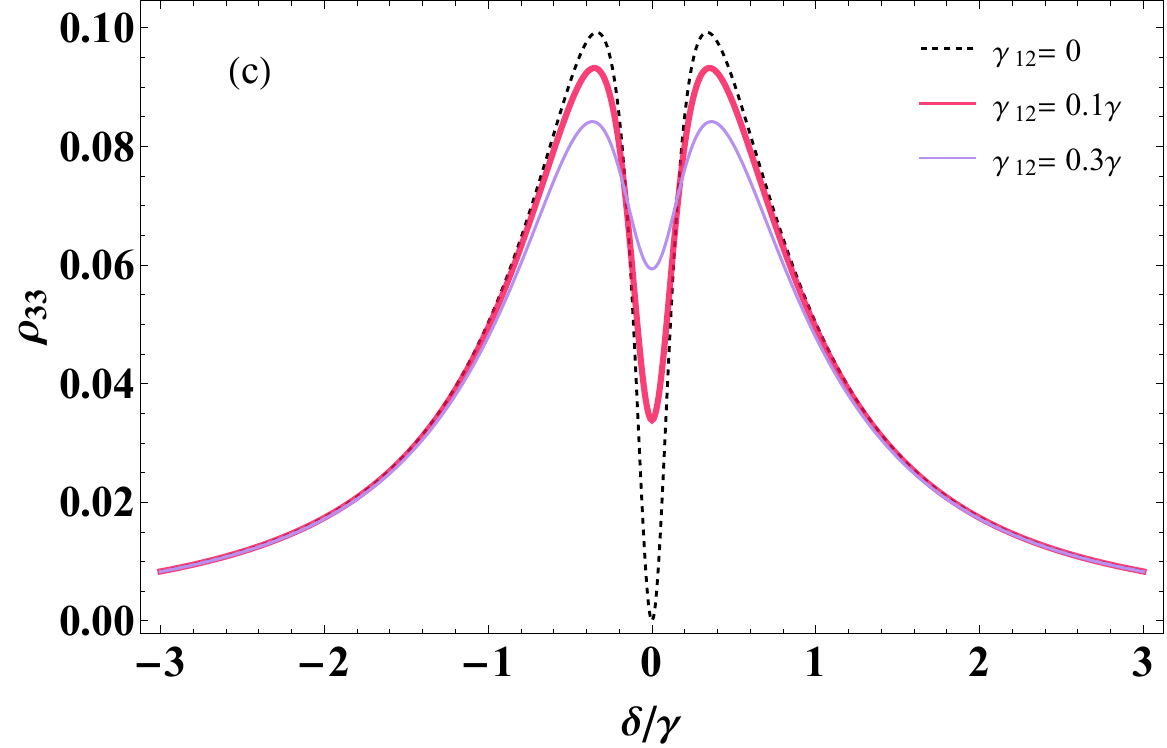}
		\includegraphics[width=8cm]{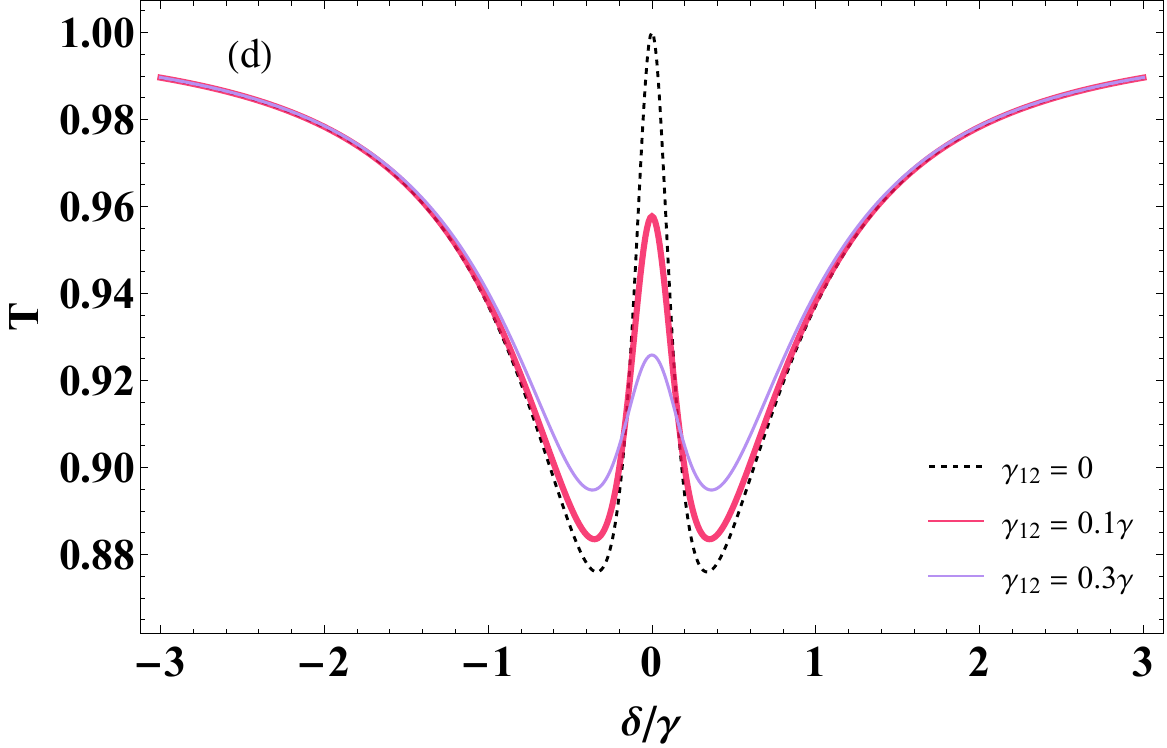}
		\caption{
			(Color online) (a) Time evolution of $\rho_{33}(\gamma t)$, and (b) time evolution of $\rho_{12}(\gamma t)$, at $\delta=0$ for three different values of $\gamma_{12}=0, 0.1\gamma, 0.3\gamma$. (c) The steady state of $\rho_{33}$, and (d) the probe transmission $T$ for three different values of $\gamma_{12}=0, 0.1\gamma, 0.3\gamma$. The Rabi frequencies have been considered as $\Omega_{c(p)}=0.4\gamma$, and the coupling laser detuning is $\Delta_c=0$.
		}
		\label{figDyphasing}
	\end{figure*}
	
	In Fig.\ref{figDynamicsDarkState} we have shown the time dependent behavior of the populations of the three energy levels as well as the real part of coherence $\rho_{12}(t)$ in the absence of noises and broadening mechanisms for the case that the Rabi frequencies of the lasers are fixed at $\Omega_{c}=\Omega_{p}=0.4\gamma$ while the coupling laser detuning and the two photon detuning have been adjusted to zero ($\Delta_c=0, \delta=0$). Besides, it has been assumed that the transition $|1\rangle\to|2\rangle$ has zero dephasing rate ($\gamma_{21}=0$). Based on our calculation, $\Im[\rho_{12}(t)]=0$ at all of the times and other coherences have zero steady state values. For this reason, we have not plotted them in Fig.\ref{figDynamicsDarkState}. As is seen from Fig.\ref{figDynamicsDarkState}, starting from an initial condition of  $ \boldsymbol{R}_0=[\, 1,\, 0,\, 0,\, \ldots\,, 0 \,]^{\text{T}}$, the system exhibits a transient behavior until it reaches a fixed steady state in which $\rho_{11}=\rho_{22}=1/2, \rho_{33}=0, \rho_{12}=-1/2$, while other density matrix elements are zero. It is obvious that this steady state is nothing except for the density matrix $\rho_{D}=|D\rangle\langle D|$ corresponding to the pure dark state of Eq.\eqref{DarkS}. The important point that should be reminded is that the final steady state of the system is independent of the initial state. It means that had the system started from any other initial state, it would have ended to the same dark state because the final steady state of Eq.\eqref{PerEqR-Zeroth}  is obtained from the following equation
	\begin{equation}\label{ssR0}
		\mathbf {M} \, \boldsymbol{R}^{(0)}_{ss} = 0,
	\end{equation}
	which is independent of the initial state. Here $\boldsymbol{R}^{(0)}_{ss}$ is the system steady state in the absence of stochastic noises.
	
	One of the most destructive broadening mechanisms that destroys the dark state effectively is the collisions between atoms that leads to the dephasing of the transition $|2\rangle\to|1\rangle$ which is the only decoherence mechanism related to the metastable level $|2\rangle$. We have already shown that in the absence of dephasing mechanism, i.e., for $\gamma_{12}=0$, the atoms reaches ideally to the dark state in the steady state at $\delta=0$. Now, we show how the dark state loses its coherence in the presence of the dephasing mechanism. For this purpose, In Figs.\ref{figDyphasing}(a) and \ref{figDyphasing}(b) we have shown the time dependent behavior of $\rho_{33}(t)$ and $\Re[\rho_{12}(t)]$ for three different dephasing rates $\gamma_{12}=0, 0.1\gamma, 0.3\gamma$ while other parameters are the same as those of Fig.\ref{figDynamicsDarkState}. As is seen the ideal dark state that is generated for $\gamma_{12}=0$ at $\delta=0$ is destroyed by increasing the value of dephasing rate because the population of the upper state is increased (Fig.\ref{figDyphasing}(a)) while the coherence of the transition $|2\rangle\to|1\rangle$, i.e., $|\rho_{12}|$, is decreased (Fig.\ref{figDyphasing}(b)).
	
	\begin{figure*}[!t]
		\centering
		\includegraphics[width=8cm]{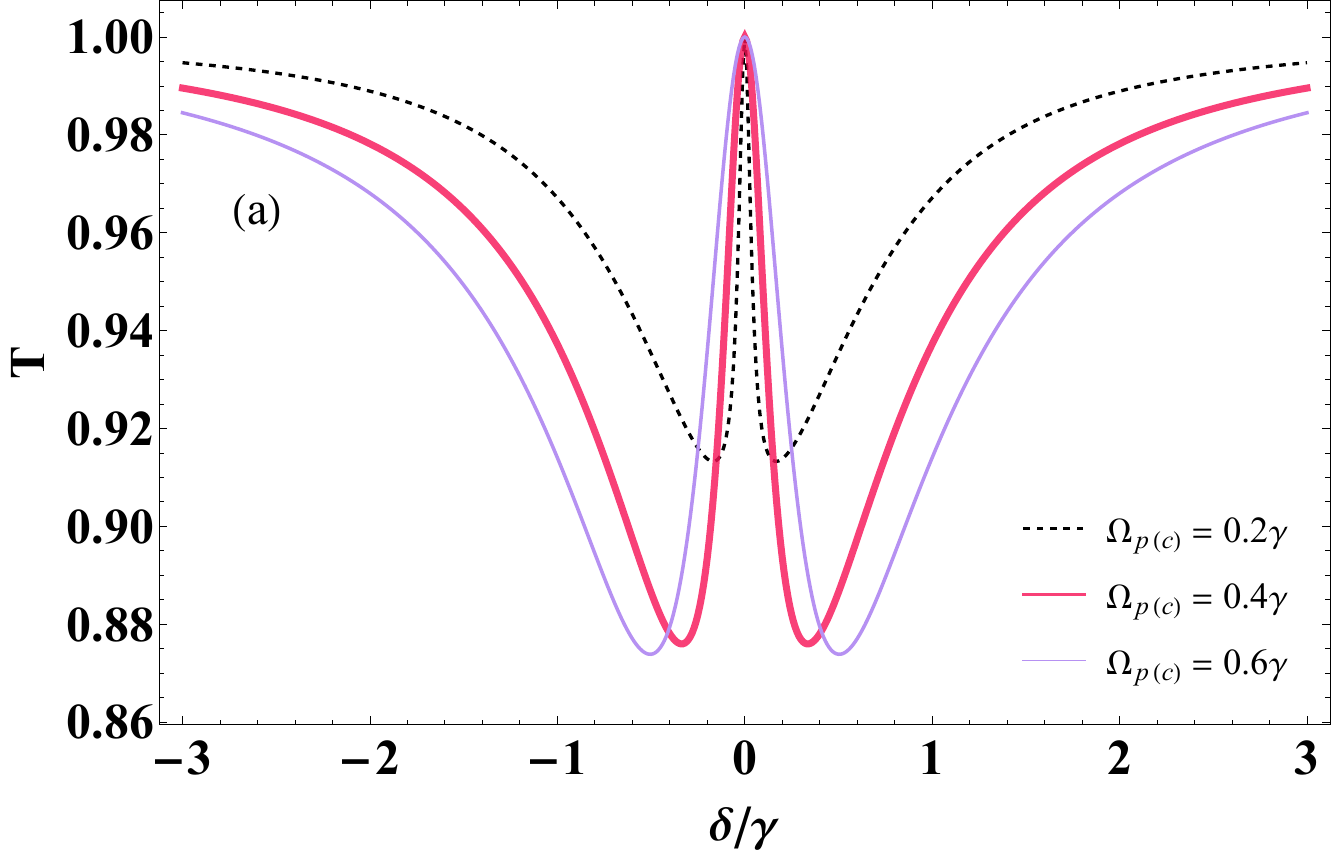}
		\includegraphics[width=8cm]{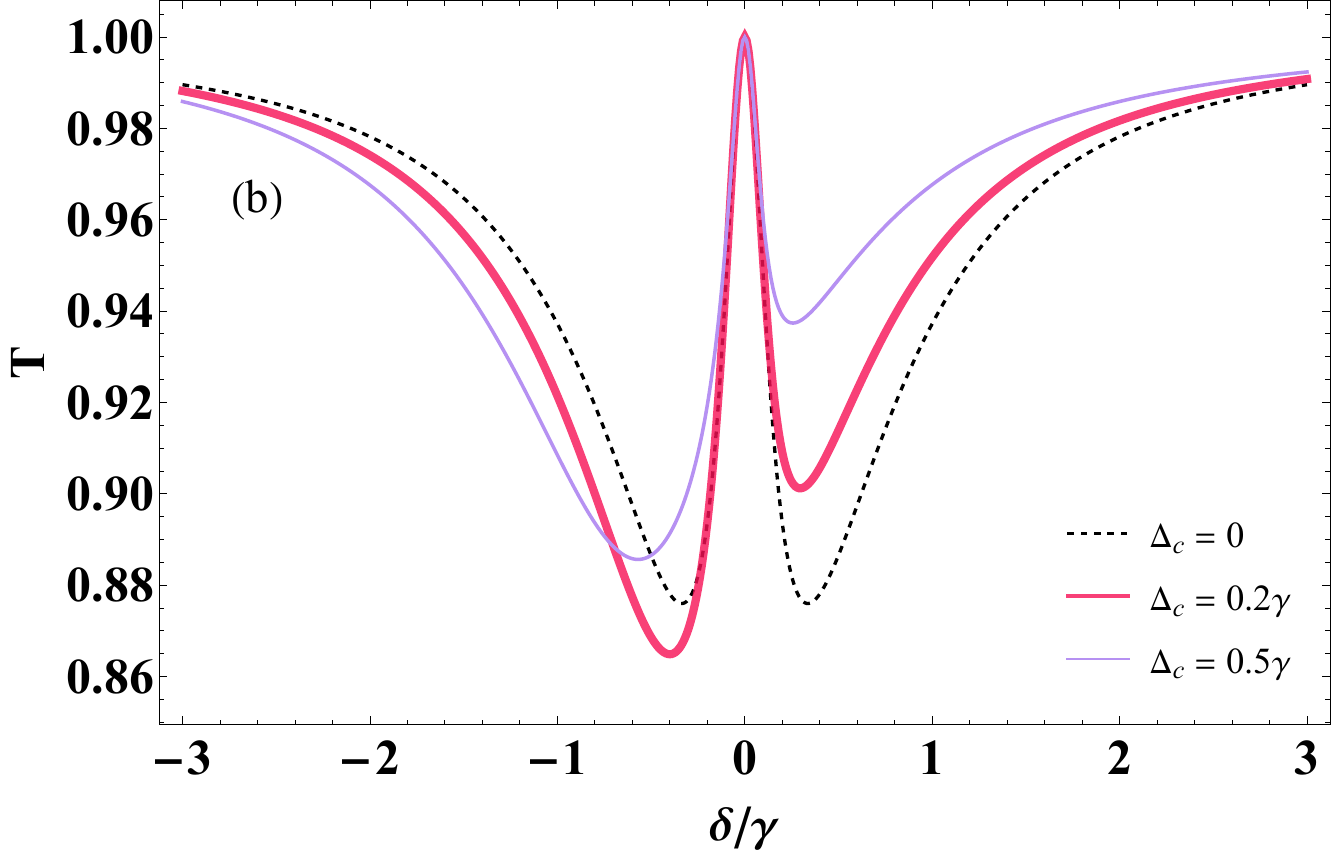}
		\includegraphics[width=8cm]{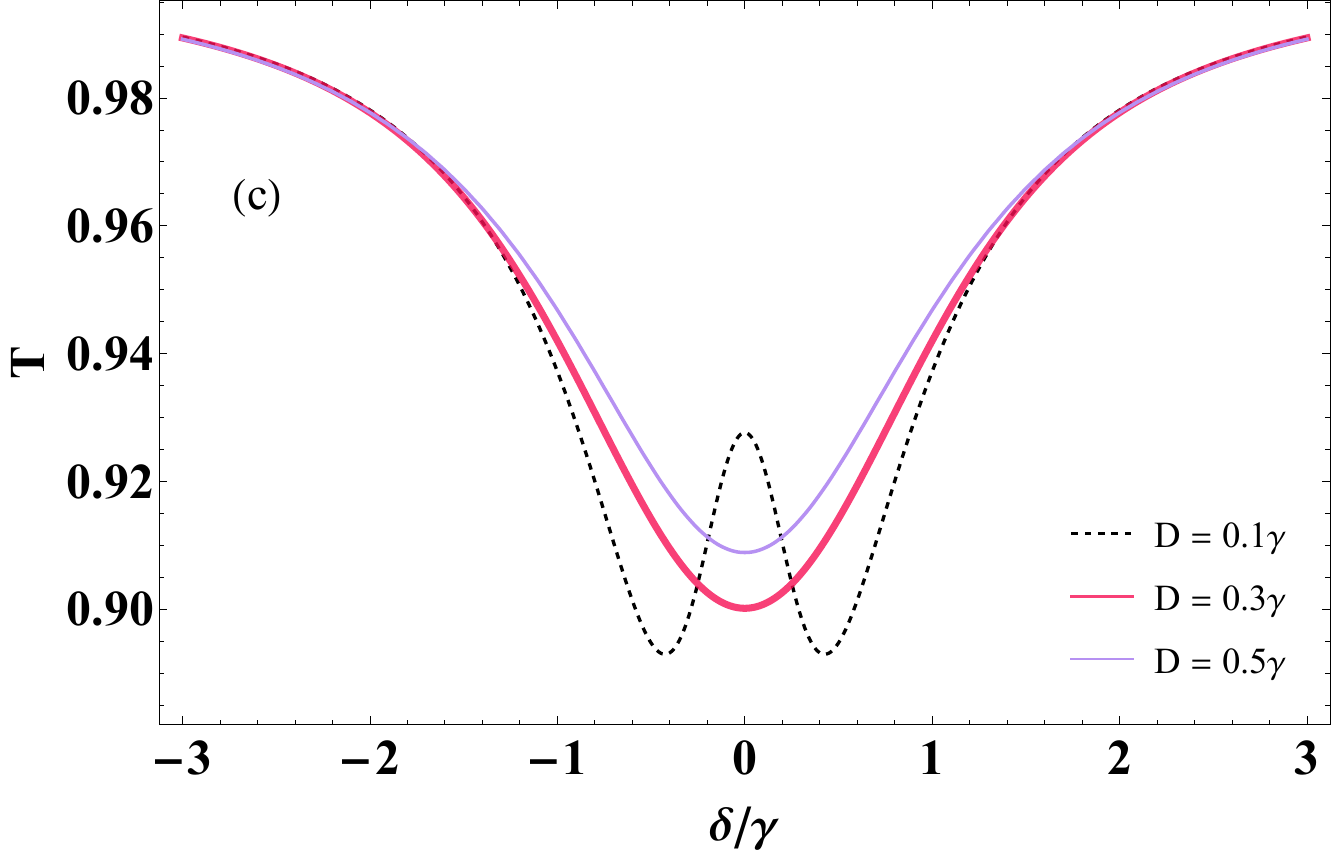}
		\includegraphics[width=8cm]{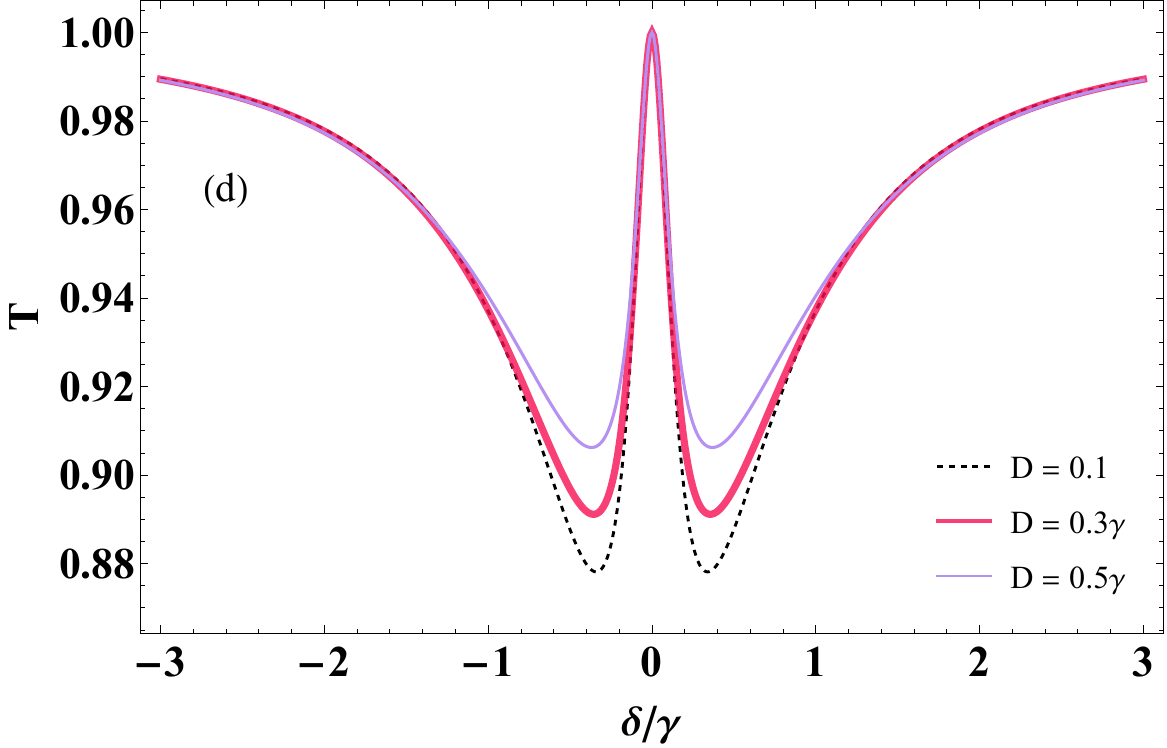}
		\caption{
			(Color online) The probe transmission versus $\delta/\gamma$ in the absence of any noises and dephasing ($\gamma_{12}=0$), (a) for three different values of Rabi frequencies $\Omega_{c(p)}=0.2\gamma, 0.4\gamma, 0.6\gamma$ at $\Delta_c=0$, and (b) for three different values $\Delta_c= 0, 0.2\gamma, 0.5\gamma$ at $\Omega_{c(p)}=0.4\gamma$. (c) and (d) for three different values of Doppler width $D=0.1\gamma, 0.3\gamma, 0.5\gamma$ for $\Omega_{c(p)}=0.4\gamma$ with $\Delta_c=0$. In (c) the two beams are copropagating and in (d) they are counterpropagating.
		}
		\label{figRabiDcDoppler}
	\end{figure*}
	
	In Figs.\ref{figDyphasing}(c), the steady state values of $\rho_{33}$ has been plotted versus the normalized two photon detuning $\delta/\gamma$ for three different dephasing rates $\gamma_{12}=0, 0.1\gamma, 0.3\gamma$ while other parameters are the same as those of Fig.\ref{figDynamicsDarkState}. Here, the steady state of the system has been calculated through solving the algebraic equation of Eq.\eqref{ssR0}. As is seen from Fig.\ref{figDyphasing}(c), there is a narrow dip (a transparency window) due to the zero normal mode of the system and two peaks corresponding to the two nonzero normal modes. As has been explained in Ref.\cite{Imamoglu} these normal modes have frequencies which are given by
	\begin{equation}\label{Normalf}
		\omega_{\pm}=\frac{1}{2}\Delta_c \pm \frac{1}{2}\sqrt{\Delta_c^2+\Omega_{p}^2+\Omega_{c}^2}.
	\end{equation}
	In the present case, where $\Delta_c=0$ and $\Omega_{p(c)}=0.4\gamma$, the peaks of normal modes appear at $\delta=\omega_{\pm}=\pm\sqrt{2}\Omega_c/2\approx\pm 0.3\gamma$. Our results show that the dephasing makes both the height of peaks of the normal modes and the depth of transparency window decrease simultaneously which leads to a broadening of the transparency window without changing the normal frequencies of Eq.\eqref{Normalf}. As is seen from Fig.\ref{figDyphasing}(c), for $\gamma_{12}=0.3\gamma$ the depth of transparency window becomes very small while its width increases.

	In order to show more clearly the destructive effect of dephasing on the performance of an atomic clock based on CPT, we have plotted in Fig.\ref{figDyphasing}(d) the transmission coefficient of the probe laser at the steady state which is defined as
	\begin{equation}
		T=1-\Im[\rho_{31}]_{ss},
	\end{equation}
	for three different dephasing rates $\gamma_{12}=0, 0.1\gamma, 0.3\gamma$. In the absence of dephasing ($\gamma_{12}=0$) there exists a sharp peak with $100\%$ transmission at $\delta=0$ corresponding to the narrow transparency window of Fig.\ref{figDyphasing}(c) (the so-called CPT transmission resonance line) which is the direct consequence of the formation of the Dark state. It means that in the absence of any kind of noises, broadening mechanism, and decoherence, the probe laser has a maximum transmission in a narrow interval around $\delta=0$. However, by increasing the value of dephasing rate, it is seen that the height of CPT transmission resonance is decreased from top and bottom so that for $\gamma_{12}=0.3\gamma$ it becomes very small which is due to the disappearance of the Dark state.

	It should also be noted that the Rabi frequency of the coupling laser is the origin of another kind of broadening mechanism called the power broadening. In this context, there are two important regimes. For $\Omega_{c}<\gamma/4$ the system lies in the weak coupling regime leading to the manifestation of the electromagnetically induced transparency (EIT) phenomenon \cite{Nori} while for $\Omega_{c}>\gamma/4$ the system lies in the strong coupling regime leading to the manifestation of the Autler-Townes splitting \cite{Nori}. Similar phenomena named optomechanically induced transparency (OMIT) and normal mode splitting (NMS) occur in the standard optomechanical systems \cite{OMIT Huang,OMIT Weis} as well as in the hybrid optomechanical systems consisting of Bose-Einstein condensates \cite{mik1,Askari1} in weak and strong coupling regimes, respectively.
	
	In order to see this matter more clearly, we have shown in Fig.\ref{figRabiDcDoppler}(a) the transmission coefficient of the probe laser versus $\delta/\gamma$ at $\Delta_c=0$ for three different values of Rabi frequencies $\Omega_{c(p)}=0.2\gamma, 0.4\gamma, 0.6\gamma$ in the absence of any noises and dephasing ($\gamma_{12}=0$). In all of the plots of probe transmission presented in Fig.\ref{figRabiDcDoppler}, we have used Eq.\eqref{ssR0} to calculate the steady state of the system in order to obtain $[\rho_{31}]_{ss}$. As is seen for $\Omega_{c(p)}=0.2\gamma$ (the black dashed curve in Fig.\ref{figRabiDcDoppler}(a)), which is lower than the threshold of $\gamma/4$, the system is in the EIT regime and therefore the transparency window is pretty narrow and consequently the CPT transmission resonance line is very sharp. However, by increasing the Rabi frequencies above the threshold of $\gamma/4$, where the system lies in the Autler-Townes splitting regime, the distance between the two normal modes is increased based on Eq.\eqref{Normalf} and consequently the CPT transmission resonance line is broadened. Obviously the larger the Rabi frequencies, the farther the distance of the normal modes (the two dips) from each other and the wider the CPT transmission resonance line. That is why in the atomic clocks based on the CPT phenomenon it is preferable to decrease the Rabie frequencies so far that the signal is maintained against the noise

	On the other hand, in order to show the effect of coupling laser detuning on the CPT line, we have plotted the probe transmission versus $\delta/\gamma$ for three different values of coupling detuning $\Delta_c=0, 0.2\gamma, 0.5\gamma$ for a fixed value of Rabi frquency $\Omega_{c(p)}=0.4\gamma$ in the absence of dephasing ($\gamma_{12}=0$) and other noises in Fig.~\ref{figRabiDcDoppler}(b). As is seen, any deviation in the coupling detuning destroys the symmetric shape of the CPT transmission line and makes it asymmetric. This effect can be used to fine tune the coupling frequency to the resonance condition of $\Delta_c=0$ which corresponds to the most symmetric CPT transmission resonance line \cite{Belcher2009}.
	
	The other broadening mechanism that destroys the coherence of the dark state in warm gaseous atoms is the Doppler broadening which is an inhomogeneous broadening mechanism \cite{EberlyTwolevel}. In Figs.\ref{figRabiDcDoppler}(c) and \ref{figRabiDcDoppler}(d) we have shown the effect of Doppler broadening on the probe transmission for the situation where both the Rabi frequencies are $\Omega_{c(p)}=0.4\gamma$ with $\Delta_c=0$ for three different values of Doppler width $D=0.1\gamma, 0.3\gamma, 0.5\gamma$ in the absence of dephasing and other noises. In Fig.\ref{figRabiDcDoppler}(c) it has been assumed that the two coupling and probe beams are copropagating, while in Fig.\ref{figRabiDcDoppler}(d) it has been assumed that the two beams are counterpropagating. As is seen clearly, increase of the Doppler width over $D=0.3\gamma$ makes the CPT transmission resonance line disappear completely when the two beams are copropagating (Fig.\ref{figRabiDcDoppler}(c)) while the same values of Doppler width has a fairly weak effect on the probe transmission when the two beams are counterpropagating (Fig.\ref{figRabiDcDoppler}(d)). Therefore, in order to neutralize the effect of Doppler broadening it suffices to make an arrangement in the experimental setup so that the two coupling and probe beams propagate in the opposite directions as has been demonstrated in Fig.\ref{figSchematic}. Nevertheless, even for counterpropagating beams the height of CPT transmission resonance line decreases from bottom by increasing the Doppler width as is seen from Fig.\ref{figRabiDcDoppler} (d).

	\begin{figure*}[!t]
		\centering
		\includegraphics[width=8cm]{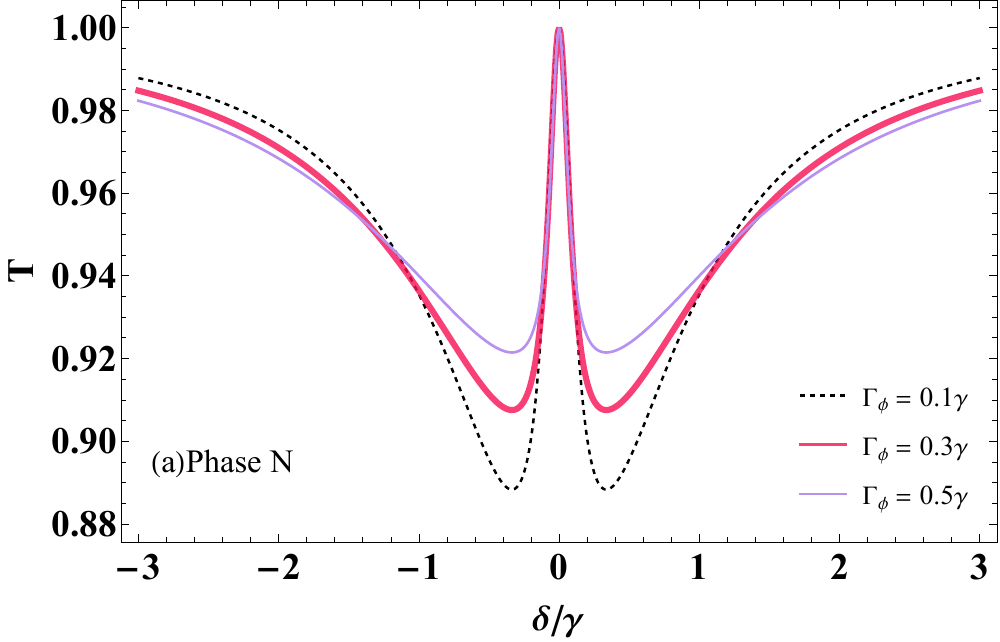}
		\includegraphics[width=8cm]{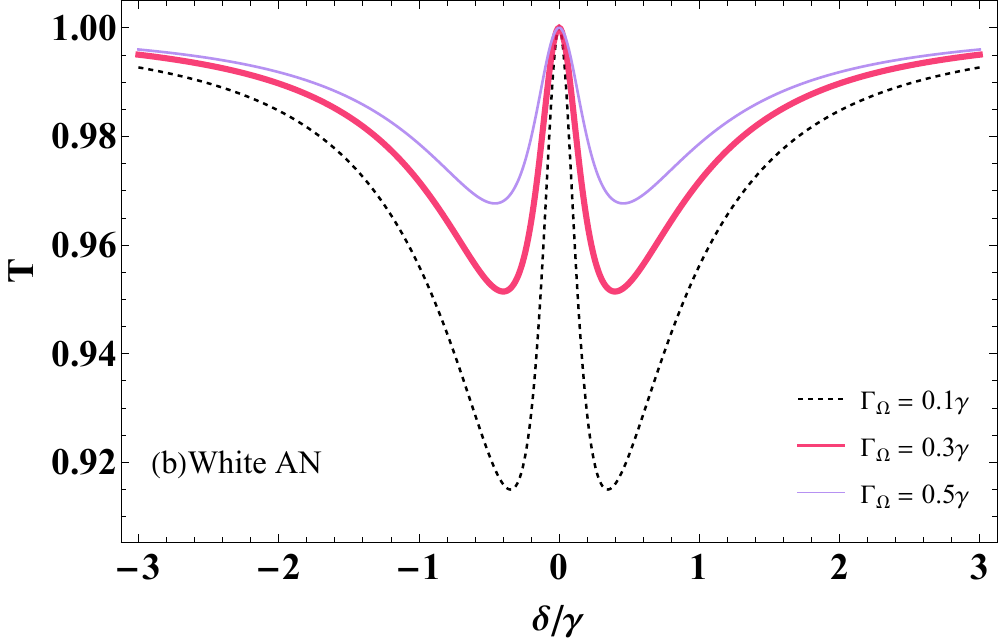}
		\includegraphics[width=8cm]{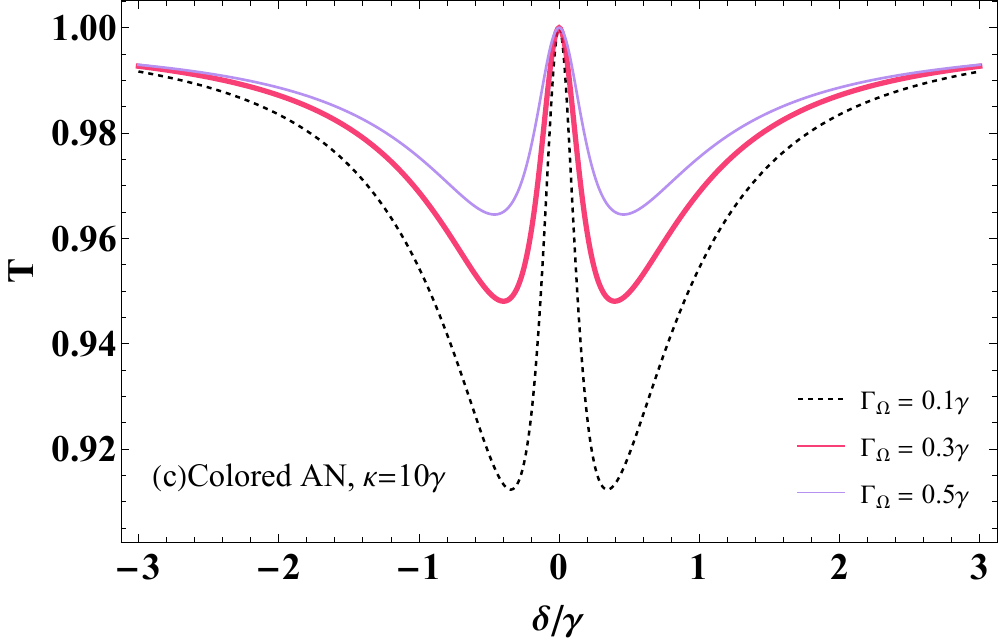}
		\includegraphics[width=8cm]{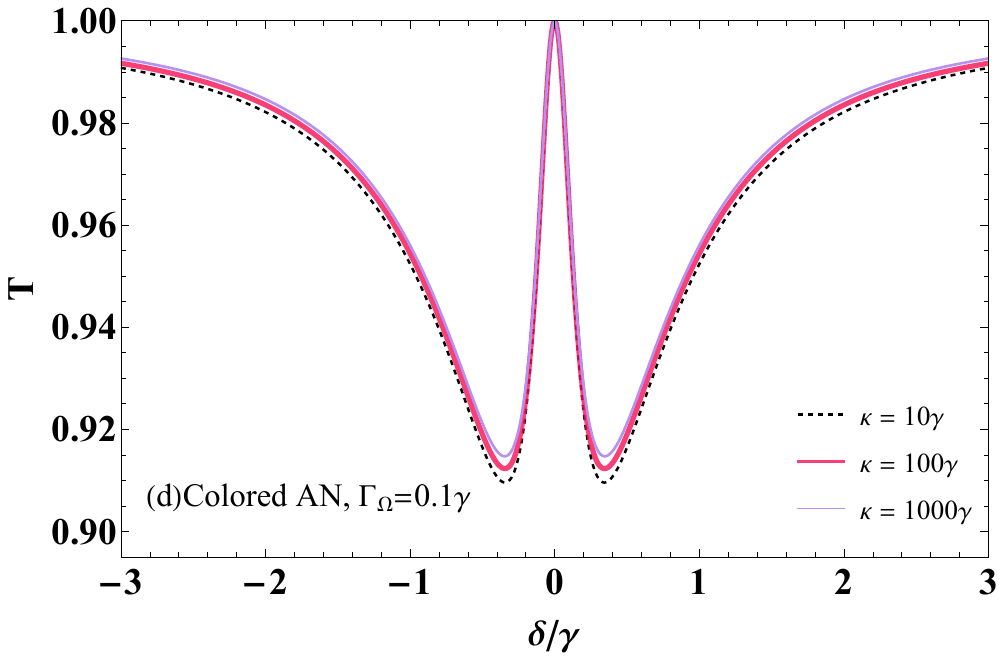}
		\caption{
			(Color online) The probe transmission versus $\delta/\gamma$ in the absence of Doppler broadening and dephasing at $\Delta_c=0$ with $\Omega_{c(p)}=0.4\gamma$, (a) for three different values of $\Gamma_{\phi}=0.1\gamma, 0.3\gamma, 0.5\gamma$ with $\Gamma_{\Omega}=0$ , and (b) for three different values of $\Gamma_{\Omega}=0.1\gamma, 0.3\gamma, 0.5\gamma$ with $\Gamma_{\phi}=0$. In (c) the amplitude fluctuation has been considered as a colored noise with a bandwidth of $\kappa=10\gamma$ for three different strengths of $\Gamma_{\Omega}=0.1\gamma, 0.3\gamma, 0.5\gamma$, while in (d) it has been considered as a colored noise with a strength of $\Gamma_{\Omega}=0.1\kappa$ for three different bandwidths of $\kappa=10\gamma, 100\gamma, 100\gamma$, in the absence of the phase noise ($\Gamma_{\phi}=0$).
		}
		\label{figNoise}
	\end{figure*}

	Therefore, in comparison to the dephasing mechanism, which makes the height of CPT transmission resonance line decrease from top and bottom simultaneously, the destroying effect of Doppler broadening in the counterpropagating configuration is very slighter. The height decrease of CPT transmission resonance line from bottom, which is due to the increase of the off-resonance probe transmission, causes the contrast between on-resonance signal and off-resonance background light to decrease which can make the signal detection become harder in the experimental setup of a CPT-based atomic clock. So, it is preferable to laser cool and trap the atoms in order to decrease the Doppler width as much as possible. Nevertheless, even for heavy alkali atoms like Rubidium 87 whose $D_1$ line ($5 ^{2}S_{1/2} \rightarrow 5 ^{2}P_{1/2}$) has a natural line width of $\gamma\approx36.13$ MHz and a Doppler width of $D\approx 37\gamma\approx 1.3$ GHz  \cite{Rb87Steck} at room temperature, the CPT transmission resonance line is still detectable although the off resonance transmission reaches to $99\%$ based on our calculations.

	It should be noted that the contrast is defined as the deference between CPT transmission resonance signal and off-resonance transmission background light, divided by CPT transmission resonance. In order to express it in percent, it should be multiplied by 100, so that it can be written as \cite{Shah2007}
	\begin{equation}\label{contrast}
		C=\frac{h_{max}^{CPT}-h_{min}^{CPT}}{h_{max}^{CPT}}\times 100,
	\end{equation}
	where $h_{max}^{CPT}$ is the height of CPT transmission resonance line (or the probe transmission exactly at $\delta=0$), and $h_{min}^{CPT}$ is the amount of off-resonance probe transmission near $\delta=0$ where its value becomes minimum.
	
	In Fig.\ref{figNoise} we have shown the effects of stochastic noises due to phase and amplitude fluctuations on the CPT resonance line in the absence of Doppler broadening and dephasing at $\Delta_c=0$ with $\Omega_{c(p)}=0.4\gamma$. In Fig.\ref{figNoise} (a) the effect of phase noise (Phase N) on the probe transmission has been plotted versus $\delta/\gamma$ for three strengths of $\Gamma_{\phi}=0.1\gamma, 0.3\gamma, 0.5\gamma$ in the absence of amplitude noise ($\Gamma_{\Omega}=0$), while in Fig.\ref{figNoise}(b) the effect of a white amplitude noise (White AN) on the probe transmission has been plotted versus $\delta/\gamma$ for three strengths of $\Gamma_{\Omega}=0.1\gamma, 0.3\gamma, 0.5\gamma$ in the absence of phase noise ($\Gamma_{\phi}=0$). On the other hand, in Fig.\ref{figNoise} (c) the effect of a colored amplitude noise (Colored AN) with a bandwidth of $\kappa=10\gamma$ on the probe transmission has been plotted versus $\delta/\gamma$ for three different strengths of $\Gamma_{\Omega}=0.1\gamma, 0.3\gamma, 0.5\gamma$, in the absence of phase noise ($\Gamma_{\phi}=0$), while in Fig.\ref{figNoise}(d) the effect of a colored amplitude noise (Colored AN) with a strength of $\Gamma_{\Omega}=0.1\gamma$ on the probe transmission has been plotted versus $\delta/\gamma$ for three different bandwidths of $\kappa=10\gamma, 100\gamma, 1000\gamma$, in the absence of the phase noise ($\Gamma_{\phi}=0$).

	It should be reminded that both in Fig.\ref{figNoise}(a) with $\Gamma_{\Omega}=0$, and in Fig.\ref{figNoise}(b), where the amplitude fluctuation has been considered as a white noise, Eq.\eqref{StStWhiteNoise} has been used, while in Figs.\ref{figNoise}(c) and \ref{figNoise}(d), where the amplitude fluctuation has been considered as a colored noise, Eq.\eqref{StStColoredN} has been used for calculation of the steady state solution. The effect of phase noise on the CPT resonance line depicted in Fig.\ref{figNoise}(a) is due to the modification of decoherence rates $\gamma_{31}$ and $\gamma_{32}$ as was explained in subsection \ref{subwhitenoise}. As is clearly seen from Figs.\ref{figNoise}(a) and \ref{figNoise}(b) under equal conditions, the destroying effect of amplitude noise is much more considerable than that of the phase noise. Fig.\ref{figNoise}(b) shows that for higher strengths of the amplitude noise, not only the height of the CPT transmission resonance line is considerably reduced from the bottom but also its width is increased, while for the same strengths of the phase noise there are just height reduction from the bottom with no increase in the width. It means that the laser \textit{phase noise} reduces the \textit{contrast} of the CPT resonance line without affecting its spectral width. This result which has been obtained through our exact approach confirms the validity of an approximate method which had been previously presented in Ref.\cite{Matveev2008}. A comparison of Fig.\ref{figNoise}(a) with Fig.\ref{figRabiDcDoppler}(d) shows that although the destroying effect of phase noise on the height reduction in CPT transmission resonance line is greater than that of the Doppler broadening in the counter propagating configuration under similar conditions, their effects are very similar. This is while, the effect of amplitude noise is much more destructive. 
	
	\begin{figure*}[!t]
		\centering
		\includegraphics[width=11cm]{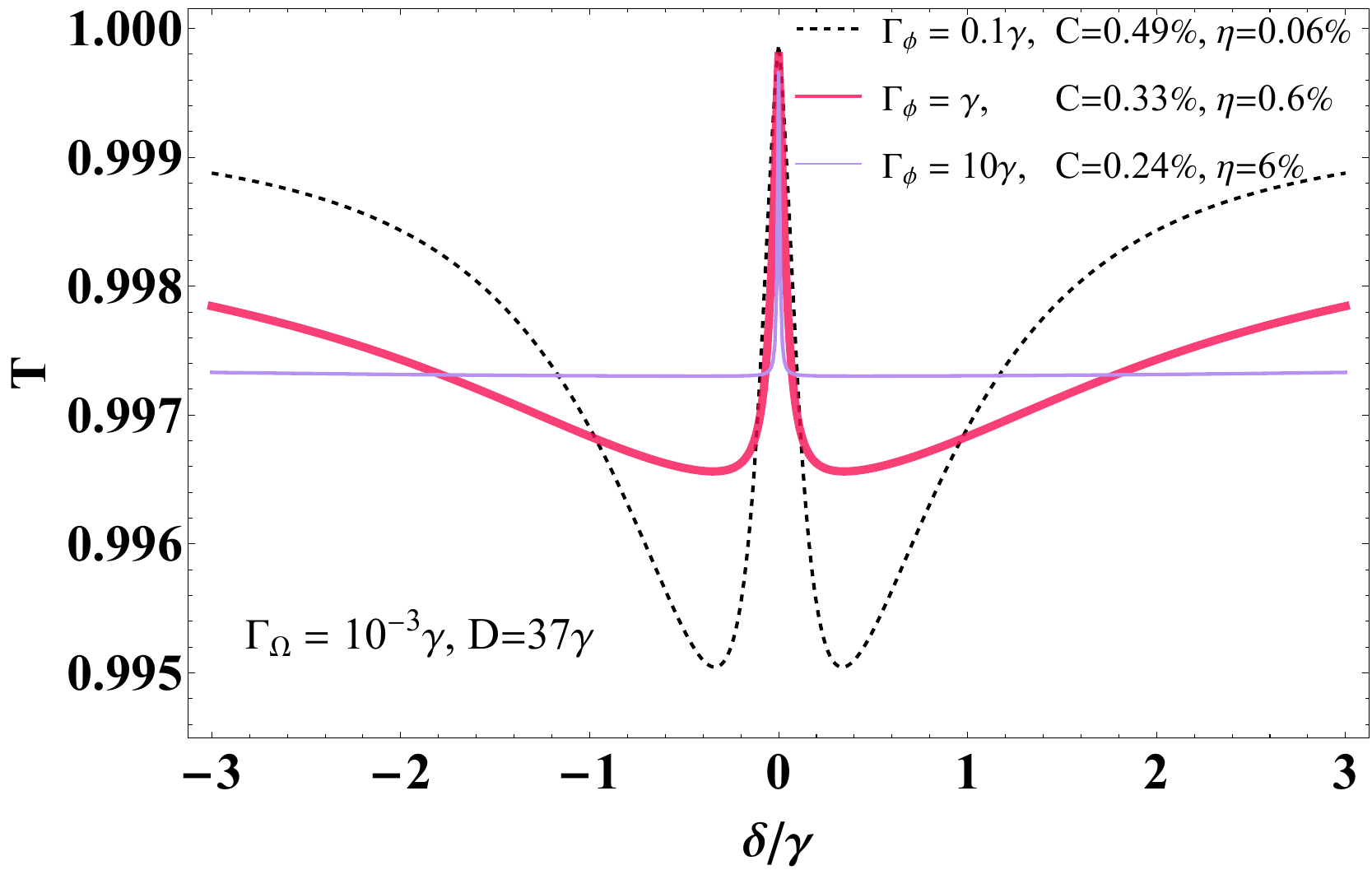}
		\caption{
			Calculation of contrast of CPT transmission resonance line defined by Eq.\eqref{contrast} for a room temperature atomic gas of $^{87}$Rb while it is driven by a laser with amplitude noise strength of $\Gamma_{\Omega}=10^{-3}\gamma=36$ KHz for different values of phase noise strength. It has been assumed that the coupling beam is on resonance with the hyperfine transition $F=2\to F'=2$ of the $D_1$ line of $^{87}$Rb ($\Delta_c=0$) while the probe beam frequency is swept across the hyperfine transition $F=1\to F'=2$ and the coupling and probe beams are counterpropagating with Rabi frequencies $\Omega_{c(p)} = 0.4 \gamma$. For the $D_1$ line ($5 ^{2}S_{1/2} \rightarrow 5 ^{2}P_{1/2}$) of $^{87}$Rb with a natural line width of $\gamma\approx36.13$ MHz, there exists a Doppler width of  $D\approx 37\gamma\approx 1.3$ GHz at room temperature  while the dephasing rate of meta stable ground state $5 ^{2}S_{1/2} $ with $F=2$ can be considered as $\gamma_{12}=10^{-3}\gamma$. The parameter $\eta$ has been defined by Eq.\eqref{eta0}
		}
		\label{figclock}
	\end{figure*}
	
	On the other hand, if the amplitude fluctuation is considered as a colored noise with a bandwidth of $\kappa=10\gamma$, as is seen from Fig.\ref{figNoise}(c), a very similar behavior like that of Fig.\ref{figNoise}(b) is observed. In order to see the effect of colored amplitude noise with different bandwidths with a fixed value of $\Gamma_{\Omega}=0.1\gamma$, in Fig.\ref{figNoise}(d) it is seen that even for a bandwidth as low as $\kappa=10\gamma$ the difference between the colored (black dashed curve in Fig.\ref{figNoise}(d)) and white (black dashed curve in Fig.\ref{figNoise}(b)) amplitude noises under the same conditions  is negligible. As is expected, the larger the bandwidth $\kappa$ of a colored amplitude noise, the closer its behavior to that of the white noise. These results, shows clearly that the white noise model of the amplitude fluctuation is a good approximation.
	
	\subsection{Experimental discussion}
	
	In the end, we would like to show how the contrast of CPT transmission resonance line (Eq.\eqref{contrast} ) in a real CPT-based atomic clock, equipped with a glass cell containing a gas of $^{87}$Rb atoms at room temperature, is affected by the Doppler broadening as well as laser noises simultaneously. For this purpose, it should be noted that the $D_1$ line ($5 ^{2}S_{1/2} \rightarrow 5 ^{2}P_{1/2}$) of $^{87}$Rb with a central transition wavelength of $\lambda=794.978 851 156(23)$nm has a natural line width of $\gamma\approx36.13$ MHz \cite{Rb87Steck}. For an atomic gas of $^{87}$Rb at room temperature, there exists a Doppler width of $D\approx 37\gamma\approx 1.3$ GHz  while the dephasing rate of meta stable ground state $5 ^{2}S_{1/2} $ with $F=2$ (corresponding to the level $|2\rangle$ in our model) can be considered as $\gamma_{12}=10^{-3}\gamma$ \cite{Belcher2009}. If the coupling beam is on resonance with the hyperfine transition $F=2\to F'=2$ of the $D_1$ line ($\Delta_c=0$) while the probe beam frequency is swept across the hyperfine transition $F=1\to F'=2$, a CPT transmission resonance line like those of Fig.\ref{figclock} becomes observable if the Rabi frequencies of lasers are fixed at $\Omega_{c(p)}=0.4\gamma$ and the coupling and probe beams are counterpropagating like the situation demonstrated in Fig.\ref{figSchematic}.

	In Fig.\ref{figclock}, we have plotted the probe transmission versus $\delta/\gamma$ for three different values of phase noise strength ($\Gamma_{\phi}$) for a fixed value of amplitude noise strength of $\Gamma_{\Omega}=10^{-3}\gamma=36$kHz. It should be noted that the parameter $\eta$ demonstrated in Fig.\ref{figclock} has been defined as the ratio of the first term of Eq.\eqref{whitepowerEd} to the second term (at $\omega=\omega_d$). In order ro express it in percent, it should be multiplied by 100 so that it is obtained as
	\begin{equation}\label{eta0}
		\eta=\frac{\Gamma_{\phi}\Gamma_{\Omega}}{\Omega_{c}}\times 100.
	\end{equation}
	For a real laser whose electric field is given by Eq.\eqref{Epcdef} with determined values of $\Gamma_{\Omega}$, $\Gamma_{\phi}$, and $\Omega_{d}$, the parameter $\eta$ shows the ratio of the power corresponding to the fluctuating part, i.e., $\delta\mathcal{E}(t)\exp(-i(\omega_d t +\phi(t)))$, to the mean power of laser corresponding to $\mathcal{E}_d\exp(-i(\omega_d t +\phi(t)))$. The lower is the parameter $\eta$ the more stable is the laser.

	As is seen from Fig.\ref{figclock}, the contrast is decreased from $C=0.49\%$ for a laser with $\eta=0.06\%$ corresponding to $\Gamma_{\phi}=0.1\gamma\approx 3.6$MHz, to $C=0.24\%$ for a laser with $\eta=6\%$ corresponding to $\Gamma_{\phi}=10\gamma\approx 360$MHz. It should be emphasized that although such predicted contrasts seems to be very low, but even a contrast as low as $0.1\%$ is enough for the performance of a CPT-based atomic clock in practice \cite{Shah2007}. Based on our calculations, if the phase noise strength reaches to $\Gamma_{\phi}=2D=72\gamma\approx 2.6$GHz (corresponding to a laser with $\eta$=45\%) the contrast decreased to $0.08\%$ and consequently the CPT transmission resonance line is no longer detectable.

	Finally, it should be noted that based on our calculation if the atoms are laser cooled and trapped so that their effective temperature is reduced to sub-Kelvin ($\approx 0.1$mK), corresponding to the Doppler width of $D=0.02\gamma\approx 0.72$MHz, the contrast can be increased to the considerable amount of $C=10.9\%$ for a laser with $\Gamma_{\phi}=0.1\gamma\approx 3.6$MHz, $\Gamma_{\Omega}=10^{-3}\gamma\approx 36$ KHz (corresponding to $\eta=0.06\%$). It shows why laser cooling and trapping of atoms are so important for the performance enhancement of atomic clocks. Compare this result with the room temperature case demonstrated by the black dashed curve in Fig.\ref{figclock} with a contrast of $C=0.49\%$ having similar noises. 
	

	\section{Summary and conclusion}\label{secSummary}
	In this paper, we have developed a mathematical formalism to investigate classical multiplicative stochastic processes based on the GF approach which has already been used in QFT and quantum many body physics. In this approach, the GF of an interacting quantum system is obtained from the noninteracting (free of noise) one through an infinite perturbative series which is shown that it converges to an exact closed form when all orders of perturbation up to infinity are taken into account. Inspired from this idea, we have shown that such an approach can be generalized to the classical multiplicative stochastic processes in which the classical noises enter into the dynamical equations in a multiplicative way as driving stochastic forces.

	We have applied the GF approach to the multiplicative processes to study a phenomenon in atomic physics, called coherent population trapping (CPT), in which the atoms are driven by two beams extracted from the same laser source so that the atomic levels form a $\Lambda$-shape structure. Since in real experimental setups the lasers have phase and amplitude fluctuations, we have modeled the pumping fields as classical stochastic variables so that the system dynamics derived from the master equation takes the form of a classical homogeneous Langevin equation in which the classical laser phase and amplitude noises enter in a multiplicative way. It should be emphasized that although we have applied the above mentioned approach to a quantum system, it can also be applied to any dynamical system (no matter being quantum or classical) with multiplicative stochastic noises. We have shown that the system of homogeneous multiplicative SDE can be solved using the perturbation theory so that it transforms to a system of linear inhomogeneous Langevin equations in every order of perturbation which can be solved based on the solutions to the previous order. Then, by taking expectation values, the interacting GF is derived as an infinite perturbative series whose each order is obtained in terms of noninteracting GF. If the stochastic variables are white (delta correlated) noises, the infinite series corresponding to the interacting GF converges exactly to a closed form so that the exact time evolution of the system as well as its exact steady state are derived from it. Nevertheless, the presented formalism has the advantage to be extended to dynamical systems with multiplicative colored noises. We have shown that if the bandwidth of the colored noise is sufficiently larger than the system damping rate, the infinite series corresponding to the interacting GF can be approximated by the closed form. In this way, the dynamics of the system and its steady state can be simply calculated again through the closed form of interacting GF with a good approximation.
	
	Using the noninteracting (noise free) GF, we have firstly shown that in the absence of any kind of noises or broadening mechanisms the system evolves in time until it reaches to a steady state known as the dark state under the two photon resonance condition. Besides, the collision and power broadenings are studied using the noninteracting GF. Then, using the closed form of interacting GF, we have studied the effects of classical laser phase and amplitude noises on the CPT phenomenon. Specifically, in order to show more clearly the destructive effect of them on the performance of an atomic clock based on CPT, we have studied their effects on the transmission coefficient of the probe laser which scans the frequencies around the transition $|3\rangle\to|1\rangle$. The most destructive broadening mechanisms that destroys the coherence of dark state is the collisions between atoms that leads to the dephasing of the transition $|2\rangle\to|1\rangle$ which is the only decoherence mechanism related to the metastable level $|2\rangle$. Our results show that the dephasing makes the height of CPT transmission resonance line decrease both from top and bottom, and also broadens its width. In the second place (after the dephasing mechanism) lies the amplitude noise which both reduces the height of CPT transmission resonance line from the bottom and increases the width of it, while the phase noise lies in the third place where just reduces the height of CPT resonance line from bottom without changing the width.
	
	On the other hand, the Doppler broadening is another kind of broadening mechanism, called inhomogeneous broadening, that is very important in warm gaseous atoms. In the phenomenon of CPT, where the atoms are exposed to the radiation of two beams, the Doppler broadening becomes very destroying if the beams are copropagating. We have shown that in the absence of dephasing, for a Doppler width of $D=0.3\gamma$ the CPT transmission line disappears completely if the two probe and coupling beams are copropagating, while in the absence of Doppler broadening, under equal conditions (for the same values of $\Delta_c$ and $\Omega_{c(p)}$) for a dephasing rate of $\gamma_{12}=0.3\gamma$ the CPT resonance line still survives although its height becomes very small and its width increases. It shows that for copropagating beams the destroying effect of Doppler broadening is even more severe than that of collision broadening (dephasing mechanism). However, if the beams are counterpropagating the destroying effect of Doppler broadening is considerably reduced so that for $D=0.3\gamma$ the CPT transmission resonance line is changed very slightly. The effect of Doppler broadening in the counterpropagating configuration is similar but slighter than that of phase noise.

	Finally, it should be reminded that the height decrease of CPT transmission resonance line from bottom, which is due to the increase of the off-resonance probe transmission, causes the contrast between on-resonance signal and off-resonance background light to decrease which can make the signal detection become harder in the experimental setup of an atomic clock based on the CPT. So, it is preferable to laser cool and trap the atoms in order to decrease the Doppler width as much as possible. Nevertheless, even for heavy alkali atoms like $^{87}$Rb whose $D_1$ line ($5 ^{2}S_{1/2} \rightarrow 5 ^{2}P_{1/2}$) has a natural line width of $\gamma\approx36.13$ MHz, and a Doppler width of $D\approx 37\gamma\approx 1.3$ GHz at room temperature, the CPT transmission resonance line is still detectable with a contrast in the interval $0.2\%<C<0.5\%$ based on our calculations with $\Omega_{c(p)}=0.4\gamma$, $\Gamma_{\Omega}=10^{-3}\gamma$, and $0.1\gamma<\Gamma_{\phi}<10\gamma$, corresponding to a laser with $0.06\%<\eta<6\%$. It confirms why an atomic clock based on CPT phenomenon can function even for warm gaseous atoms.

	\appendix
	\section{Spectral analysis of stochastic electric fields having amplitude and phase fluctuations}\label{AppA}
	The statistics of phase fluctuation is described by the Wiener-Levy phase diffusion model with the following Langevin equation
	\begin{equation}\label{phiDot}
		\frac{d \phi(t)}{dt} = \dot{\phi}(t)
	\end{equation}	
	where $\dot{\phi}(t)$ is considered as a Gaussian white noise \cite{Papoulis} with zero mean $\langle \dot{\phi}(t) \rangle = 0$, and the autocorrelation function of
	\begin{equation}\label{Cphifot}
		C_{\phi}(t-t')=\langle \dot{\phi}(t) \, \dot{\phi}(t') \rangle = 2\Gamma_\phi \, \delta(t - t'),
	\end{equation}
	where $\Gamma_\phi$ is the intensity of $\dot{\phi}(t)$ with the dimension of frequency. On the other hand, the amplitude fluctuation is assumed to be described by a Gaussian stochastic process with the zero mean $\langle \delta\mathcal{E}(t)\rangle = 0$ and autocorrelation function of $\langle \delta\mathcal{E}(t)\, \delta\mathcal{E}(t') \rangle = C_{\mathcal{E}}(t - t')$, which can be considered as white or colored noise depending on the form of $ C_{\mathcal{E}}(t - t')$. Note that both $\phi(t)$ and $\delta\mathcal{E}(t)$ are real stochastic variables. Moreover, the fluctuations in the phase and amplitude, described by independent stochastic variables, are statistically uncorrelated and have independent Gaussian distribution functions so that $\langle \dot{\phi}(t) \, \delta\mathcal{E}(t') \rangle = 0$. Such fluctuations make the fields be stochastic variables whose statistical properties have been determine by their autocorrelation function
	\begin{equation}\label{CEtt}
		C_{E_d}(t-t')=\langle E_d(t)E^{\ast}_d(t')\rangle,
	\end{equation}
	where $d=c$ or $p$ indicates the coupling or probe beam. On substitution from Eq.\eqref{Epcdef} in the autocorrelation function of Eq.\eqref{CEtt}, it can be seen that
	\begin{equation}\label{expandCEtt}
		C_{E_d}(t-t')=\big(\mathcal{E}_0^2 + C_{\mathcal{E}}(t-t')\big)\langle e^{-i\phi(t)+i\phi(t')}\rangle e^{-i\omega_d (t-t')}.
	\end{equation}
	Now, by integrating the Langevin equation of Eq.\eqref{phiDot} it can be seen that the stochastic phase is obtained as
	\begin{eqnarray}\label{phit-t}
		\phi(t)=\phi(0)+\int_{0}^{t}\dot{\phi}(t')dt'.
	\end{eqnarray}
	It shows that the stochastic variable $\phi(t)$, which is formed by a constant $\phi(0)$ plus a sum of a large number of statistically independent stochastic variables $\dot\phi(t')$, has a Gaussian distribution based on the central limit theorem \cite{Papoulis}. Based on Eq.\eqref{phit-t}, the stochastic variable $\phi(t)-\phi(0)$ has the following second moment
	\begin{equation}\label{secmomphit0}
		\langle\big(\phi(t)-\phi(0)\big)^2\rangle=\int_{0}^{t}dt'\int_{0}^{t}dt''\langle\dot{\phi}(t')\dot{\phi}(t'')\rangle=2\Gamma_{\phi}t.
	\end{equation}
	Since $\phi(t)$ is a Gaussian stochastic variable, it can be shown by the Wick's theorem that \cite{Papoulis}
	\begin{equation}\label{expphi}
		\langle e^{-i\phi(t)+i\phi(t')}\rangle=e^{-\frac{1}{2}\langle[\phi(t')-\phi(t)]^2\rangle}
	\end{equation}
	Furthermore, from Eq.\eqref{phit-t} it can be seen that
	\begin{equation}\label{secmomphttp}
		\langle[\phi(t')-\phi(t)]^2\rangle=\int_{t}^{t'}ds\int_{t}^{t'}ds'\langle\dot\phi(s)-\dot\phi(s')\rangle=2\Gamma_{\phi}|t'-t|
	\end{equation}
	where the delta correlated form of the autocorrelation of Eq.\eqref{Cphifot} has been used in both of Eqs.\eqref{secmomphttp} and \eqref{secmomphit0}. On substitution of Eqs.\eqref{secmomphttp} and \eqref{expphi} in Eq.\eqref{expandCEtt}, it can be seen that the electric field autocorrelation function obtained as 
	\begin{equation}\label{simpexpandCEtt}
		C_{E_d}(\tau)=\big(\mathcal{E}_0^2 + C_{\mathcal{E}}(\tau)\big) e^{-i\omega_d\tau-\Gamma_{\phi}|\tau|},
	\end{equation}
	where it has been assumed that $\tau=t-t'$. In the following we study the autocorrelation function of Eq.\eqref{simpexpandCEtt} in two different cases where in the former the amplitude fluctuations is assumed to be a white noise while in the latter it is considered as a Lorentzian colored noise.
	
	\subsection{Case I: white amplitude noise}
	If the stochastic variable $\delta\mathcal{E}(t)$ is a white (delta correlated) noise, then $\delta\Omega(t)$ is also a white noise with the following autocorrelation
	\begin{equation}\label{CWwhit}
		C_\Omega(\tau)=2\Gamma_\Omega \delta(\tau),
	\end{equation}
	because of the relation $C_{\Omega}(t - t') = \frac{d^2}{\hbar^2} C_{\mathcal{E}}(t - t')$. Here, $\Gamma_\Omega$ has been considered as the intensity of the stochastic variable $\delta\Omega(t)$ with the dimension of frequency. Therefore, if the amplitude fluctuation is a white noise, then the field autocorrelation function of Eq.\eqref{simpexpandCEtt} is read as
	\begin{equation}
		C_{E_d}(\tau)=\Big(\mathcal{E}_0^2 + \frac{2\hbar^2}{d^2}\Gamma_{\Omega}\delta(\tau)\Big) e^{-i\omega_d\tau-\Gamma_{\phi}|\tau|},
	\end{equation}
	and consequently the power spectrum corresponding to the field $E_d(t)$ which is defined as \cite{Papoulis}
	\begin{equation}\label{powerspectrumdef}
		S_{E_d}(\omega)=\int_{-\infty}^{+\infty} C_{E_d} (\tau)e^{-i\omega\tau}d\tau, 
	\end{equation}
	is calculated as
	\begin{equation}\label{whitepowerEd}
		S_{E_d}(\omega)=\frac{2\hbar^2}{d^2}\Gamma_\Omega+\frac{2\Gamma_{\phi}\mathcal{E}_0^{2}}{(\omega-\omega_d)^2+\Gamma_\phi^2}.
	\end{equation}
	As is seen, if the amplitude fluctuation is a white noise, then the power spectrum of $E_d(t)$ has a uniform contribution due to the amplitude noise plus a Lorentzian contribution due to the phase noise with a spectral width of $\Delta\omega=\Gamma_{\phi}$.
	
	\subsection{case II: colored amplitude noise}
	If the stochastic variable $\delta\mathcal{E}(t)$ is a Lorentzian colored noise, then $\delta\Omega(t)$ is also a Lorentzian colored noise whose autocorrelation function can be considered as
	\begin{equation}\label{CWcolor}
		C_\Omega(\tau)=\kappa\Gamma_{\Omega} e^{-\kappa|\tau|},
	\end{equation}
	where $\Gamma_{\Omega}$ is the strength of the stochastic variable $\delta\Omega(t)$ and $\kappa$ is the bandwidth. So, if $\delta\mathcal{E}(t)$ is a Lorentzian colored noise, then the field autocorrelation function of Eq.\eqref{simpexpandCEtt} is read as
	\begin{equation}
		C_{E_d}(\tau)=\mathcal{E}_0^2 e^{-i\omega_d\tau-\Gamma_{\phi}|\tau|} + \frac{\hbar^2}{d^2}\kappa\Gamma_{\Omega} e^{-i\omega_d\tau-(\Gamma_{\phi}+\kappa)|\tau|},
	\end{equation}
	and the corresponding power spectrum is read as
	\begin{equation}\label{colorpowerEd}
		S_{E_d}(\omega)=\frac{2\hbar^2}{d^2}\frac{\kappa\Gamma_\Omega(\Gamma_{\phi}+\kappa)}{(\omega-\omega_d)^2+(\Gamma_{\phi}+\kappa)^2}+\frac{2\Gamma_{\phi}\mathcal{E}_0^{2}}{(\omega-\omega_d)^2+\Gamma_\phi^2},
	\end{equation}
	As is seen, if the amplitude fluctuation is a Lorentzian colored noise, then the power spectrum of $E_d(t)$ has two Lorentzian contributions where the former corresponding to the amplitude noise has a width of $\Gamma_{\phi}+\kappa$ and the latter corresponding to the phase noise has a width of $\Gamma_\phi$. It should be reminded that if the bandwidth $\kappa$ of the amplitude colored noise is very large, i.e., in the limit of $\kappa\gg\gamma$ and $\kappa\gg\Gamma_{\phi}$, the contribution of the amplitude noise power spectrum in the first term of Eq.\eqref{colorpowerEd} goes to the first term of Eq.\eqref{whitepowerEd} for the white amplitude noise as is expected.

	\section{Derivation of the fourth order GF}\label{AppB}
	In this appendix we show how Eq.\eqref{G4_t} is derived from Eq.\eqref{G4tfirstform}. For this purpose, we substitute $\bm\Phi(t)$ from Eq.\eqref{PhiMatrixdef} into Eq.\eqref{G4tfirstform} so that the result is obtained as follows
	\begin{equation}\label{R4_1}
		\begin{aligned}
			\bm{\mathcal{G}}^{(4)}(t) = \int_0^t& dt_1\int_0^{t_1} dt_2\int_0^{t_2} dt_3\int_0^{t_3} dt_4 \, \bm{G}(t-t_1)\; \\[3pt]
			\times \Big\langle\big[&\dot\phi(t_1)\mathbf{N}+\delta\Omega(t_1)\mathbf{L}\big] 
			\bm{G}(t_1-t_2)\\[2.5pt]
			\times \big[&\dot\phi(t_2)\mathbf{N}+\delta\Omega(t_2)\mathbf{L}\big] \bm{G}(t_2-t_3) \\[3pt]
			\times \big[&\dot\phi(t_3)\mathbf{N}+\delta\Omega(t_3)\mathbf{L}\big] 
			\bm{G}(t_3-t_4) \\[2.5pt]
			\times \big[&\dot\phi(t_4)\mathbf{N}+\delta\Omega(t_4)\mathbf{L}\big]\Big\rangle\,\bm{G}(t_4).
		\end{aligned}
	\end{equation}	
	By fully expanding the integrand, 16 terms are appeared as
	\begin{widetext}
		\begin{equation}\label{R4_2}
			\begin{aligned}
				\langle \bm{\mathcal{G}}^{(4)}(t)  \rangle= \int_0^t dt_1\int_0^{t_1} dt_2\int_0^{t_2} & dt_3\int_0^{t_3} dt_4  \; \bm{G}(t-t_1) \Big\{\\[2pt]
				\quad &\langle\dot\phi(t_1)\dot\phi(t_2)\dot\phi(t_3)\dot\phi(t_4)\rangle\, \mathbf{N}\,\bm{G}(t_1-t_2)\mathbf{N}\,\bm{G}(t_2-t_3)\mathbf{N}\,\bm{G}(t_3-t_4)\mathbf{N} \\[2pt]
				\quad +\; &\langle\dot\phi(t_1)\dot\phi(t_2)\dot\phi(t_3)\,\delta\Omega(t_4)\rangle\, \mathbf{N}\,\bm{G}(t_1-t_2)\mathbf{N}\,\bm{G}(t_2-t_3)\mathbf{N}\,\bm{G}(t_3-t_4)\mathbf{L} \\[2pt]
				\quad +\; &\langle\dot\phi(t_1)\dot\phi(t_2)\,\delta\Omega(t_3)\,\dot\phi(t_4)\rangle\, \mathbf{N}\,\bm{G}(t_1-t_2)\mathbf{N}\,\bm{G}(t_2-t_3)\mathbf{L}\,\bm{G}(t_3-t_4)\mathbf{N} \\[2pt]
				\quad +\; 
				&\langle\dot\phi(t_1)\dot\phi(t_2)\,\delta\Omega(t_3)\,\delta\Omega(t_4)\rangle
				\, \mathbf{N}\,\bm{G}(t_1-t_2)\mathbf{N}\,\bm{G}(t_2-t_3)\mathbf{L}\,\bm{G}(t_3-t_4)\mathbf{L} \\[2pt]
				\quad +\; 
				&\langle\dot\phi(t_1)\,\delta\Omega(t_2)\,\dot\phi(t_3)\,\dot\phi(t_4)\rangle
				\, \mathbf{N}\,\bm{G}(t_1-t_2)\mathbf{L}\,\bm{G}(t_2-t_3)\mathbf{N}\,\bm{G}(t_3-t_4)\mathbf{N} \\[2pt]
				\quad +\;
				&\langle \dot\phi(t_1)\,\delta\Omega(t_2)\,\dot\phi(t_3)\,\delta\Omega(t_4)\rangle
				\, \mathbf{N}\,\bm{G}(t_1-t_2)\mathbf{L}\,\bm{G}(t_2-t_3)\mathbf{N}\,\bm{G}(t_3-t_4)\mathbf{L} \\[2pt]
				\quad +\;
				&\langle\dot\phi(t_1)\,\delta\Omega(t_2)\,\delta\Omega(t_3)\,\dot\phi(t_4)\rangle
				\, \mathbf{N}\,\bm{G}(t_1-t_2)\mathbf{L}\,\bm{G}(t_2-t_3)\mathbf{L}\,\bm{G}(t_3-t_4)\mathbf{N}\\[2pt]
				\quad +\;
				&\langle\dot\phi(t_1)\,\delta\Omega(t_2)\,\delta\Omega(t_3)\,\delta\Omega(t_4)\rangle
				\, \mathbf{N}\,\bm{G}(t_1-t_2)\mathbf{L}\,\bm{G}(t_2-t_3)\mathbf{L}\,\bm{G}(t_3-t_4)\mathbf{L}\\[2pt]
				\quad +\;
				&\langle \delta\Omega(t_1)\,\dot\phi(t_2)\,\dot\phi(t_3)\,\dot\phi(t_4) \rangle\, \mathbf{L}\,\bm{G}(t_1-t_2)\mathbf{N}\,\bm{G}(t_2-t_3)\mathbf{N}\,\bm{G}(t_3-t_4)\mathbf{N}\\[2pt]
				\quad +\;
				&\langle \delta\Omega(t_1)\,\dot\phi(t_2)\,\dot\phi(t_3)\,\delta\Omega(t_4) \rangle\, \mathbf{L}\,\bm{G}(t_1-t_2)\mathbf{N}\,\bm{G}(t_2-t_3)\mathbf{N}\,\bm{G}(t_3-t_4)\mathbf{L}\\[2pt]
				\quad +\; 
				&\langle\delta\Omega(t_1)\,\dot\phi(t_2)\,\delta\Omega(t_3)\,\dot\phi(t_4)\rangle
				\, \mathbf{L}\,\bm{G}(t_1-t_2)\mathbf{N}\,\bm{G}(t_2-t_3)\mathbf{L}\,\bm{G}(t_3-t_4)\mathbf{N} \\[2pt]
				\quad +\;
				&\langle \delta\Omega(t_1)\dot\phi(t_2)\delta\Omega(t_3)\delta\Omega(t_4)\rangle
				\,\mathbf{L}\,\bm{G}(t_1-t_2)\mathbf{N}\,\bm{G}(t_2-t_3)\mathbf{L}\,\bm{G}(t_3-t_4)\mathbf{L} \\[2pt]
				\quad +\;
				&\langle\delta\Omega(t_1)\delta\Omega(t_2)\dot\phi(t_3)\dot\phi(t_4) \rangle\, \mathbf{L}\,\bm{G}(t_1-t_2)\mathbf{L}\,\bm{G}(t_2-t_3)\mathbf{N}\,\bm{G}(t_3-t_4)\mathbf{N} \\[2pt]
				\quad +\;
				&\langle\delta\Omega(t_1)\delta\Omega(t_2)\dot\phi(t_3)\delta\Omega(t_4)\rangle
				\, \mathbf{L}\,\bm{G}(t_1-t_2)\mathbf{L}\,\bm{G}(t_2-t_3)\mathbf{N}\,\bm{G}(t_3-t_4)\mathbf{L} \\[2pt]
				\quad +\;
				&\langle\delta\Omega(t_1)\delta\Omega(t_2)\delta\Omega(t_3)\dot\phi(t_4)\rangle
				\, \mathbf{L}\,\bm{G}(t_1-t_2)\mathbf{L}\,\bm{G}(t_2-t_3)\mathbf{L}\,\bm{G}(t_3-t_4)\mathbf{N} \\[2pt]
				\quad +\;
				&\langle\delta\Omega(t_1) \delta\Omega(t_2) \delta\Omega(t_3) \delta\Omega(t_4)\rangle
				\, \mathbf{L}\,\bm{G}(t_1-t_2)\mathbf{L}\,\bm{G}(t_2-t_3)\mathbf{L}\,\bm{G}(t_3-t_4)\mathbf{L}\Big\}\,\bm{G}(t_4).
			\end{aligned}
		\end{equation}
	\end{widetext}
	
	As is seen from Eq.\eqref{R4_2}, there is a four-point correlation function in each term. Since $\dot{\phi}(t)$ and $\delta\Omega(t)$ are zero-mean Gaussian stochastic variables, each four-point function can be written as a sum of terms containing the products of all possible two-point correlation functions based on the so-called Wick's theorem in the classical theory of stochastic variables \cite{Papoulis}. However, those four-point functions containing a single phase variable or a single amplitude variable become zero because in their Wick's expansion there exist cross correlations like $\langle \dot{\phi}(t_i) \, \delta\Omega(t_j) \rangle$ which are zero  under the assumption that the amplitude and phase noises are statistically uncorrelated. For example, consider such a four-point correlation function appeared in the penultimate line in Eq.\eqref{R4_2} whose Wick's expansion is given by the following equation
	\begin{align}\label{singlephase}
		\langle \delta\Omega(t_1) \delta\Omega(t_2) \delta\Omega(t_3) \dot{\phi}(t_4) \rangle =& 
		\langle \delta\Omega(t_1) \delta\Omega(t_2) \rangle \langle \delta\Omega(t_3) \dot{\phi}(t_4) \rangle \notag\\[3pt]
		+& \langle \delta\Omega(t_1) \delta\Omega(t_3) \rangle \langle \delta\Omega(t_2) \dot{\phi}(t_4) \rangle \notag\\[3pt]
		+& \langle \delta\Omega(t_1) \dot{\phi}(t_4) \rangle \langle \delta\Omega(t_2) \delta\Omega(t_3) \rangle,
	\end{align}
	which is zero because each term on the right hand side of Eq.\eqref{singlephase} contains the cross-correlations of phase and amplitude noises. Therefore, \eqref{R4_2} is simplified as
	\begin{widetext}
		\begin{equation}\label{R4_3}
			\begin{aligned}
				\langle \bm{\mathcal{G}}^{(4)}(t)  \rangle=& \int_0^t dt_1\int_0^{t_1} dt_2\int_0^{t_2} dt_3\int_0^{t_3} dt_4\; \bm{G}(t-t_1) \Big\{\\[3pt]
				\quad &\langle\dot\phi(t_1)\dot\phi(t_2)\dot\phi(t_3)\dot\phi(t_4)\rangle\, \mathbf{N}\,\bm{G}(t_1-t_2)\mathbf{N}\,\bm{G}(t_2-t_3)\mathbf{N}\,\bm{G}(t_3-t_4)\mathbf{N} \\[3pt]
				\quad +\; 
				&\langle\dot\phi(t_1)\dot\phi(t_2)\delta\Omega(t_3)\delta\Omega(t_4)\rangle
				\, \mathbf{N}\,\bm{G}(t_1-t_2)\mathbf{N}\,\bm{G}(t_2-t_3)\mathbf{L}\,\bm{G}(t_3-t_4)\mathbf{L} \\[3pt]
				\quad +\; 
				&\langle \dot\phi(t_1)\delta\Omega(t_2)\dot\phi(t_3)\delta\Omega(t_4)\rangle
				\, \mathbf{N}\,\bm{G}(t_1-t_2)\mathbf{L}\,\bm{G}(t_2-t_3)\mathbf{N}\,\bm{G}(t_3-t_4)\mathbf{L}\\[3pt]
				\quad +\;
				&\langle\dot\phi(t_1)\delta\Omega(t_2)\delta\Omega(t_3)\dot\phi(t_4)\rangle
				\, \mathbf{N}\,\bm{G}(t_1-t_2)\mathbf{L}\,\bm{G}(t_2-t_3)\mathbf{L}\,\bm{G}(t_3-t_4)\mathbf{N}\\[3pt]
				\quad +\;
				&\langle \delta\Omega(t_1)\dot\phi(t_2)\dot\phi(t_3)\delta\Omega(t_4) \rangle\, \mathbf{L}\,\bm{G}(t_1-t_2)\mathbf{N}\,\bm{G}(t_2-t_3)\mathbf{N}\,\bm{G}(t_3-t_4)\mathbf{L}\\[3pt]
				\quad +\; 
				&\langle\delta\Omega(t_1)\dot\phi(t_2)\delta\Omega(t_3)\dot\phi(t_4)\rangle
				\, \mathbf{L}\,\bm{G}(t_1-t_2)\mathbf{N}\,\bm{G}(t_2-t_3)\mathbf{L}\,\bm{G}(t_3-t_4)\mathbf{N} \\[3pt]
				\quad +\;
				&\langle\delta\Omega(t_1)\delta\Omega(t_2)\dot\phi(t_3)\dot\phi(t_4) \rangle\, \mathbf{L}\,\bm{G}(t_1-t_2)\mathbf{L}\,\bm{G}(t_2-t_3)\mathbf{N}\,\bm{G}(t_3-t_4)\mathbf{N} \\[3pt]
				\quad +\;
				&\langle\delta\Omega(t_1)\delta\Omega(t_2)\delta\Omega(t_3)\delta\Omega(t_4)\rangle
				\, \mathbf{L}\,\bm{G}(t_1-t_2)\mathbf{L}\,\bm{G}(t_2-t_3)\mathbf{L}\,\bm{G}(t_3-t_4)\mathbf{L}\Big\}\,\bm{G}(t_4).
			\end{aligned}
		\end{equation}
	\end{widetext}
	Here, again each four-point correlation function in Eq.\eqref{R4_3} can be written as a sum of terms containing the products of all possible two-point correlations based on the Wick's theorem. However, The integration is performed over the ordered domain $t_1>t_2>t_3>t_4$, which defines a four-dimensional region in time space. It should be noted that there exist two kinds of two-point correlations inside the Wick's expansion. The first ones are noncrossing contractions as $\langle \xi(t_1)\xi(t_2)\rangle \langle \xi(t_3)\xi(t_4)\rangle$ which are compatible with the ordering structure and the second ones are crossing contractions like $\langle \xi(t_1)\xi(t_3)\rangle \langle \xi(t_2)\xi(t_4)\rangle$ or $\langle \xi(t_1)\xi(t_4)\rangle \langle \xi(t_2)\xi(t_3)\rangle$, which are incompatible with the ordering structure. If the amplitude noise is delta correlated similar to the phase noise, the crossing contractions make the integral be calculated over the boundary of the region where the noncrossing contractions are integrated. Since the boundary of the region has zero measure with respect to the region, the crossing contractions have no contribution to the integral and consequently Eq.\eqref{R4_3} is exactly reduced to the following equation
	\begin{widetext}
		\begin{equation}\label{R4_4}
			\begin{aligned}
				\langle \bm{\mathcal{G}}^{(4)}(t)  \rangle=& \int_0^t dt_1\int_0^{t_1} dt_2\int_0^{t_2} dt_3\int_0^{t_3} dt_4\; \bm{G}(t-t_1) \Big\{\\[3pt]
				\quad &\langle\dot\phi(t_1)\dot\phi(t_2)\rangle \langle\dot\phi(t_3)\dot\phi(t_4)\rangle\, \mathbf{N}\,\bm{G}(t_1-t_2)\mathbf{N}\,\bm{G}(t_2-t_3)\mathbf{N}\,\bm{G}(t_3-t_4)\mathbf{N} \\[3pt]
				\quad +\; 
				&\langle\dot\phi(t_1)\dot\phi(t_2)\rangle \langle\delta\Omega(t_3)\delta\Omega(t_4)\rangle
				\, \mathbf{N}\,\bm{G}(t_1-t_2)\mathbf{N}\,\bm{G}(t_2-t_3)\mathbf{L}\,\bm{G}(t_3-t_4)\mathbf{L} \\[3pt]
				\quad +\; 
				&\langle\delta\Omega(t_1)\delta\Omega(t_2)\rangle \langle\dot\phi(t_3)\dot\phi(t_4) \rangle\, \mathbf{L}\,\bm{G}(t_1-t_2)\mathbf{L}\,\bm{G}(t_2-t_3)\mathbf{N}\,\bm{G}(t_3-t_4)\mathbf{N} \\[3pt]
				\quad +\;
				&\langle\delta\Omega(t_1)\delta\Omega(t_2)\rangle \langle \delta\Omega(t_3) \delta\Omega(t_4)\rangle \, \mathbf{L} \, \bm{G}(t_1-t_2) \mathbf{L} \, \bm{G}(t_2-t_3) \mathbf{L} \, \bm{G}(t_3-t_4) \mathbf{L} \Big\} \, \bm{G}(t_4).
			\end{aligned}
		\end{equation}
	\end{widetext}
	
	Nevertheless, if the amplitude fluctuation is described by a colored noise of the form of Eq.~\eqref{CWcolor} with a correlation time much shorter than the characteristic relaxation timescale of the system ($\kappa\gg\gamma$), it can be shown that Eq.\eqref{R4_3} is approximately reduced to Eq.\eqref{R4_4}. In such a case where the noise correlation function is sharply peaked around equal times and decays rapidly for temporal separations larger than $\kappa^{-1}$, it behaves approximately like a Dirac delta function. As a result, the dominant contribution in the Wick expansion of four-point functions of Eq.~\eqref{R4_3} again originates from the noncrossing contractions while the crossing contractions can be neglected approximately. In this way, Eq.\eqref{R4_4} will be again valid with a good approximation.
	
	Now, by using Eqs.\eqref{SigmaPhit}-\eqref{SigmaOmeghat}, Eq.\eqref{R4_4} can be written as
	\begin{align}\label{R4_5}
		\langle \bm{\mathcal{G}}^{(4)}(t)  \rangle =& \int_0^t dt_1\int_0^{t_1} dt_2\int_0^{t_2} dt_3\int_0^{t_3} dt_4\; \bm{G}(t-t_1) \notag \\
		\times \Big\{ \,&\bm{\Sigma}_{\phi}(t_1-t_2)\,\bm{G}(t_2-t_3)\,\bm{\Sigma}_{\phi}(t_3-t_4) \notag \\
		+&\bm{\Sigma}_{\phi}(t_1-t_2)\,\bm{G}(t_2-t_3)\,\bm{\Sigma}_{\Omega}(t_3-t_4) \notag \\
		+&\bm{\Sigma}_{\Omega}(t_1-t_2)\,\bm{G}(t_2-t_3)\,\bm{\Sigma}_{\phi}(t_3-t_4) \notag \\
		+&\bm{\Sigma}_{\Omega}(t_1-t_2)\,\bm{G}(t_2-t_3)\,\bm{\Sigma}_{\Omega}(t_3-t_4)\Big\}\,\bm{G}(t_4).
	\end{align}
	
	Finally, by simplifying the above equation and using Eq.\eqref{Secordselfenergy} we obtain the following equation
	\begin{equation}\label{R4_6}
		\begin{aligned}
			\langle \bm{\mathcal{G}}^{(4)}(t)  \rangle=& \int_0^t dt_1\int_0^{t_1} dt_2\int_0^{t_2} dt_3\int_0^{t_3} dt_4\; \\
			\times \bm{G}(t&-t_1)  \bm{\Sigma}(t_1-t_2)\,\bm{G}(t_2-t_3)\,\bm{\Sigma}(t_3-t_4) \,\bm{G}(t_4).
		\end{aligned}
	\end{equation}


\begin{thebibliography}{100}
		
		\bibitem{Alzetta} G. Alzetta, A. Gozzini, M. Moi, G. Orriols,
		\href{https://doi.org/10.1007/BF02749417}{Nuovo Cimento B \textbf{36}, 5 (1976)}.
		
		\bibitem{Vanier2005} J. Vanier,
		\href{https://link.springer.com/article/10.1007/s00340-005-1905-3}{Appl. Phys. B \textbf{81}, 421 (2005)}.
		
		\bibitem{Zubairy Book} M. O. Scully, and M. S. Zubairy. \textit{Quantum optics}, (Cambridge University Press, 1997).
		
		\bibitem{Belcher2009} N. Belcher, E.E. Mikhailov, and I. Novikova, 
		\href{http://dx.doi.org/10.1119/1.3120262}{Am. J. Phys. \textbf{77}, 988 (2009)}.
		
		\bibitem{Zhong Review} W. Zhong
		\href{https://iopscience.iop.org/article/10.1088/1674-1056/23/3/030601}{Chin. Phys. B \textbf{23}, No. 3  030601 (2014)}.
		
		\bibitem{minicpt1} J. Kitching; S. Knappe, and L. Hollberg
		\href{https://doi.org/10.1063/1.1494115}{Appl. Phys. Lett. \textbf{81}, 553 (2002)}
		
		\bibitem{minicpt2} S. Knappe, V. Shah, P. D. D. Schwindt, L. Hollberg, J. Kitching, Li-Anne Liew, and J. Moreland
		\href{https://doi.org/10.1063/1.1787942}{Appl. Phys. Lett. \textbf{85}, 1460 (2004)}.
		
		\bibitem{minicpt3} S. Knappe, P.D.D. Schwindt, V. Shah, L. Hollberg, J. Kitching, L. Liew, and J. Moreland
		\href{https://doi.org/10.1364/OPEX.13.001249}{Opt. Express \textbf{13}, 1249 (2005)}.
		
		
		\bibitem{Shah2007} V. Shah, S. Knappe, L. Hollberg, and J. Kitching
		\href{https://opg.optica.org/ol/abstract.cfm?URI=ol-32-10-1244}{Opt. Lett. \textbf{32}, 1244 (2007)}.
		
		\bibitem{Carmichael} H. J. Carmichael, \textit{Statistical Methods in Quantum Optics 1 }, fourth ed. (Springer-Verlag, Berlin, Heidelberg, 1999).
		
		\bibitem{Loudon} R. Loudon, \textit{The Quantum Theory of Light}, third ed. (Oxford University Press, 2001).
		
		\bibitem{Papoulis} A. Papoulis, and S. U. Pillai, \textit{Probability, Random Variables, and Stochastic Processes}, fourth ed. (McGraw-Hill, New York, 2002).
		
		\bibitem{Dop1 Vemuri} G. Vemuri, and G. S. Agarwal,
		\href{https://journals.aps.org/pra/abstract/10.1103/PhysRevA.53.1060}{Phys. Rev. A \textbf{53}, 1060 (1996)}
		
		\bibitem{Dop2 Mompart} J. Mompart, V. Ahufinger, R Corbalan, and F. Prati
		\href{https://iopscience.iop.org/article/10.1088/1464-4266/2/3/321}{J. Opt. B: Quantum Semiclass. Opt. \textbf{2}, 359 (2000)}.
		
		\bibitem{Dop3 Fan} X. Fan, C. Liu, S. Tian, J. Li, M. Zhu, N. Cui, and S. Gong
		\href{https://www.tandfonline.com/doi/abs/10.1080/09500340408235531#}{J. Mod. Opt. \textbf{51}, No. 1, 399 (2004)}.
		
		\bibitem{Redfield} A. Redfield, \textit{Advances in Magnetic Resonance}, (Academic, New York, 1965) Vol. 1, pp.1-32.
		
		\bibitem{Kubo 1962} R. Kubo, \textit{Fluctuations, Relaxation and Resonance in Magnetic Systems}, 
		(Oliver and Boyd, Edinburgh, 1962),  pp. 23-68.
		
		\bibitem{Kubo 1963} R. Kubo, 
		\href{https://doi.org/10.1063/1.1703941}{J. Math. Phys. \textbf{4}, 174 (1963)}.
		
		\bibitem{Fox} B. F. Fox, 
		\href{http://dx.doi.org/10.1063/1.1666123}{J. Math. Phys. \textbf{13}, 1196 (1972)}.
		
		\bibitem{Wodkiewicz} K. Wodkiewicz
		\href{http://dx.doi.org/10.1063/1.523960}{J. Math. Phys. \textbf{20}, 45 (1979)}.
		
		\bibitem{Cook} R. J. Cook,
		\href{https://journals.aps.org/pra/abstract/10.1103/PhysRevA.21.268}{Phys. Rev. A \textbf{21}, 268 (1980)}
		
		\bibitem{Agarwal1976} G. S. Agarwal
		\href{https://doi.org/10.1103/PhysRevLett.37.1383}{Phys. Rev. Lett. \textbf{37}, 1383 (1976)}.
		
		\bibitem{Agarwal1990} G. Vemuri, R. Roy, and G. S. Agarwal
		\href{https://doi.org/10.1103/PhysRevA.41.2749}{Phys. Rev. A \textbf{41}, 2749 (1990)}.
		
		\bibitem{Lawande1987} S. V. Lawande, R. D'Souza, and R. R. Puri
		\href{https://doi.org/10.1103/PhysRevA.36.3228}{Phys. Rev. A \textbf{36}, 3228 (1987)}.
		
		\bibitem{Agarwal1978} G. S. Agarwal
		\href{https://doi.org/10.1103/PhysRevA.18.1490}{Phys. Rev. A \textbf{18}, 1490 (1978)}.
		
		\bibitem{Dalton1982} B. J. Dalton, and P. L. Knight
		\href{https://iopscience.iop.org/article/10.1088/0022-3700/15/21/019}{J. Phys. B: At. Mol. Phys. \textbf{15}, 3997 (1982)}.
		
		\bibitem{Swain1998} P. Zhou, and S. Swain
		\href{https://doi.org/10.1103/PhysRevA.58.4705}{Phys. Rev. A \textbf{58}, 4705 (1998)}.
		
		\bibitem{Kiely2021} A. Kiely
		\href{https://doi.org/10.1209/0295-5075/134/10001}{EPL, \textbf{134}, 10001 (2021)}.
		
		\bibitem{Franco2019} B. Gu and I. Franco,
		\href{https://doi.org/10.1063/1.5099499}{J. Chem. Phys. \textbf{151}, 014109 (2019).}
		
		\bibitem{Zeng2024} J. Zeng, G. H. Xu, W. Huang, and Y Yao
		\href{https://doi.org/10.1103/PhysRevA.110.062219}{Phys. Rev. A \textbf{110}, 062219 (2024)}.
		
		\bibitem{Danageozian} A. Danageozian, A. Miller, P. J. Barge, N. Bhusal, and J. P. Dowling
		\href{https://iopscience.iop.org/article/10.1088/1361-6455/ac7760}{J. Phys. B: At. Mol. Opt. Phys. \textbf{55} 155503 (2022)}.
		
		
		\bibitem{Jiang2023} X. Jiang, J. Scott, Mark Friesen, and M. Saffman
		\href{https://doi.org/10.1103/PhysRevA.107.042611}{Phys. Rev. A \textbf{107}, 042611 (2023)}.
		
		\bibitem{Flensberg} H. Bruus and K. Flensberg, \textit{Many-Body Quantum Theory in Condensed Matter Physics: An Introduction} (Oxford University Press, Oxford, 2004).
		
		\bibitem{Rb87Steck} D. A. Steck, \textit{Rubidium 87 D Line Data},
		\href{https://steck.us/alkalidata/}{Oregon Center for Optics and Department of Physics, University of Oregon, revision 2.1.4, 2010.}
		
		\bibitem{Imamoglu} M. Fleischhauer, A. Imamoglu, J. P. Marangos,
		\href{}{Rev. Mod. Phys.\textbf{77},633, (2005)}
		
		\bibitem{Nori} B. Peng, S. K. Ozdemir, W. Chen, F. Nori, and L. Yang, 
		\href{https://doi.org/10.1038/ncomms6082}{Nat. Commun. \textbf{5}, 5082 (2014)}.
		
		\bibitem{OMIT Huang} G. S. Agarwal and S. Huang, 
		\href{https://doi.org/10.1103/PhysRevA.81.041803}{Phys. Rev. A. \textbf{81}, 041803 (2010)}. 
		
		\bibitem{OMIT Weis} S. Weis, R. Rivière, S. Deléglise, E. Gavartin, O. Arcizet, A. Schliesser, T. J. Kippenberg, 
		\href{https://doi.org/10.1126/science.1195596}{Science \textbf{330}, 1520 (2010)}.
		
		\bibitem{mik1} H. Mikaeili, A. Dalafi, M. Ghanaatshoar, and B. Askari,
		\href{https://doi.org/10.1038/s41598-022-08250-9}{Sci. Rep. \textbf{12}, 4428 (2022)}
		
		\bibitem{Askari1} B. Askari, and A. Dalafi
		\href{https://doi.org/10.1088/1751-8121/ac40e2}{J. Phys. A: Math. Theor. \textbf{55}, 035301 (2022)}
		
		\bibitem{EberlyTwolevel} L. Allen, ans J. H. Eberly, \textit{Optical Resonance and Two-Level Atoms}, (Dover Publications, INC, New York, 1987)
		
		\bibitem{Matveev2008} A. N. Matveev, A. V. Sokolov, A. V. Akimov, V. N. Sorokin, A. Yu. Samokotin, and N. N. Kolachevsky
		\href{https://doi.org/10.3103/S1068335608050059}{Bulletin of the Lebedev Physics Institute, \textbf{35}, No. 5, pp. 148–155 (Allerton Press, Inc., 2008)}.
		
		
		\bibitem{breuer2002theory}
		H.-P. Breuer and F. Petruccione, 
		\textit{The Theory of Open Quantum Systems} 
		(Oxford University Press, Oxford, 2002).
		
		
		
	\end{thebibliography}
	
\end{document}